\documentclass[traditabstract]{aa}
\usepackage{txfonts}
\usepackage{graphicx}
\usepackage{color}

\begin{document}

\title{
The chemical composition of a regular halo globular cluster:\\ NGC 5897\thanks{This paper includes data gathered with the 6.5 meter Magellan Telescopes located at Las Campanas Observatory, Chile.}
}

\author{Andreas Koch\inst{1} 
  \and Andrew McWilliam\inst{2}
  }
  
\authorrunning{A. Koch \& A. McWilliam}
\titlerunning{Chemical abundances in NGC 5897}
\offprints{A. Koch;  \email{akoch@lsw.uni-heidelberg.de; andy@obs.carnegiescience.edu}}

\institute{
Zentrum f\"ur Astronomie der Universit\"at Heidelberg, Landessternwarte, K\"onigstuhl 12, 69117 Heidelberg, Germany
  \and Carnegie Observatories, 813 Santa Barbara St., Pasadena, CA 91101, USA
    }
\date{}
\abstract {
We report  for the first time on the chemical composition of the halo cluster NGC~5897 (R$_{\odot}$=12.5 kpc), based on 
chemical abundance ratios for 27 $\alpha$-, iron-peak, and neutron-capture elements in seven red giants. 
From our high-resolution, high signal-to-noise 
spectra obtained with the Magellan/MIKE spectrograph, 
we find a mean iron abundance from the neutral species of [Fe/H]=$-2.04\pm0.01$ (stat.) $\pm0.15$ (sys.), which is more metal-poor than implied by 
previous photometric and low-resolution spectroscopic studies. 
NGC~5897 is $\alpha$-enhanced (to 0.34$\pm$0.01 dex) and shows Fe-peak element ratios typical of other (metal-poor) halo globular clusters (GCs)  
 with no overall, significant abundance spreads in iron nor in any other heavy element. %
Like other GCs,  NGC~5897 shows a clear Na-O anti-correlation, where we find a prominent primordial population of stars with enhanced O abundances and 
$\sim$Solar Na/Fe ratios, while 
two stars are Na-rich, providing chemical proof of the presence of multiple populations in this cluster. 
Comparison of the heavy element abundances with the Solar-scaled values and  the metal poor GC M15 from the literature confirms that 
NGC~5897 has experienced only little contribution from $s$-process nucleosynthesis. 
One star of the first generation stands out in that it shows very low La and Eu abundances.
Overall, NGC~5897 is a well-behaved GC showing archetypical correlations and element-patterns, with little room for surprises in our data. 
We suggest that its lower metallicity could explain the unusually long periods of RR Lyr that were found in NGC~5897.
}
\keywords{Stars: abundances -- Galaxy: abundances -- Galaxy: evolution -- Galaxy: halo -- globular clusters: individual: NGC~5897}
\maketitle 
%
%
%
%
%
%
\section{Introduction}
Globular clusters (GCs) are amongst the oldest stellar systems in the Universe and purvey important information  
on the earliest evolutionary stages of the Galaxy. 
As such, they have received much attention over the past years as they also harbor a multitude of fascinating properties. 
Long believed to be {\em the} classical example of simple stellar populations,  it is now accepted that GCs 
consist of multiple stellar populations that are evident in their color-magnitude diagrams (CMDs; e.g., Piotto 2009, and references therein) and their light chemical 
element distributions (e.g., Gratton et al. 2012, and references therein. These are a direct consequence of the presence of at least two generations of stars, differing in age and chemical abundances,  
that formed within a few 100 Myr of each other.  Stars of the first generations (either fast rotating massive stars, Decressin et al. 2007, or intermediate-mass Asymptotic Giant Branch [AGB] stars, 
e.g., D'Ercole et al. 2008) pollute the Interstellar Medium (ISM) with the products of p-capture reactions, out of which the dominant, second stellar generation is born. 
This enrichment scenario leaves the $\alpha$- and Fe-peak  elements unaltered (as is indeed observed in GCs), while producing the characteristic 
light-element variations such as the Na-O and Mg-Al anti-correlations and CN and CH bimodalities. 

Considering the broad variety in the GCs'  properties (horizontal branch [HB] morphology, central concentration, 
environment, to name a few), it is then striking that several key features are found throughout every single GC studied to date. For instance, 
several authors have now resorted to {\em defining} a GC as an object that shows a Na-O correlation (Carretta et al. 2010). 
In this work,  we will add the poorly-studied inner halo GC NGC~5897 to the family of typical GCs by measuring in-depth for the first time its chemical composition. 
This will reveal if also NGC~5897 exhibits the canonical chemical  abundance variations and how these could relate to the factors of, e.g.,  environment.

NGC~5897 is a metal-poor GC in the Galactic halo, located at a moderate distance from the Sun  (R$_{\odot}$=12.5 kpc; R$_{\rm GC}$=7 kpc; Harris 2006 [2010 edition].).  
It shows a predominantly blue HB with a  morphology that has been labeled ``normal'' considering its low metallicity (e.g., Clement \& Rowe 2001; Fig.~1). 
However, some of  its other properties still defy a satisfying explanation:

In a photometric study, Testa et al. (2001) analysed NGC~5897 and three other GCs, which were all confirmed to be coeval 
and to have similar metallicity. 
Thus it was concluded that there must exist a very prominent,  {additional} parameter besides age and metallicity governing 
the differences in their  HB morphologies. This parameter was taken to be ``environment'', represented, e.g., by the concentration 
or central densities of such stellar systems ({Fusi-Pecci et al. 1993}). 
In this context, NGC~5897 has a  very low concentration (half-light  and tidal radii are  2.1$\arcmin$ and 10.1$\arcmin$, respectively);  
its concentration parameter (c=$\log\,[r_t/r_c]$) is 0.86 and, in fact,  only 15\% of the entire known Milky Way  GCs are less densely
concentrated, where 60\% of those are within the inner halo at R$_{\rm GC}<20$ kpc (Harris 1996). 
{Further, possible "third parameters" comprise the Helium abundance of the clusters (Gratton et al. 2010).}

It is very surprising that, given the wealth of photometric data for this GC, no consensus as to its metallicity is in sight, yet: the
current values  for [Fe/H] in the literature, obtained from photometry or  low-resolution spectroscopy, range from 
$-1.68$ (Zinn \& West 1984) to $-2.09$ dex (Kraft \& Ivans 2003), which is by far a 
larger uncertainty  than the possible accuracy achievable with present-day spectroscopic methods. 
Such an ignorance would also seem to  prohibit an accurate
 assessment  even of the simplest parameters governing the HB morphology. 

NGC~5897  is known to harbour several variable stars, mainly RR~Lyrae. Their periods were measured to be longer than 0.6 days, which is not unusual for a low-metallicity cluster.
Puzzling is, however, the unparalleled, long mean period of the RR$ab$ variables of $\left<P_{ab}\right> >$ 0.8 d 
 (e.g., Wehlau 1990; Clement et al. 2001; Clement \& Rowe 2001), which could indicate even lower metallicities than the mean of this GC. 
 
This work is organized as follows: In \S2 we describe our target selection and data gathering; the abundance and error analyses are laid out in detail in 
\S\S3,4. All our results are presented and interpreted in \S5 before summarizing our findings in \S6.
\section{Observations and data reduction}
\subsection{Target selection}
Eight red giant member candidates were selected for observations 
from the photometric study of Testa et al. (2001) such as to cover the brightest giants while avoiding the centrally crowded regions of this cluster\footnote{Despite
its loose concentration, NGC 5897's large distance renders the most central regions too dense for single-star slit spectroscopy.}. 
In particular, our targets are located 2.0$\arcmin$ to 5.1$\arcmin$ ($\sim$1--2.5 half-light radii) from the cluster center. 
We ensured that all targets are sufficiently unaffected by crowding (Stetson et al. 2003) and 
in only one case there is an estimated $\sim$0.9\% flux contribution from near-by sources, while it is essentially zero for all other targets. 
The respective CMD including our target stars is shown in Fig.~1. 
%
%
\begin{figure}[htb]
\centering
\includegraphics[width=1\hsize]{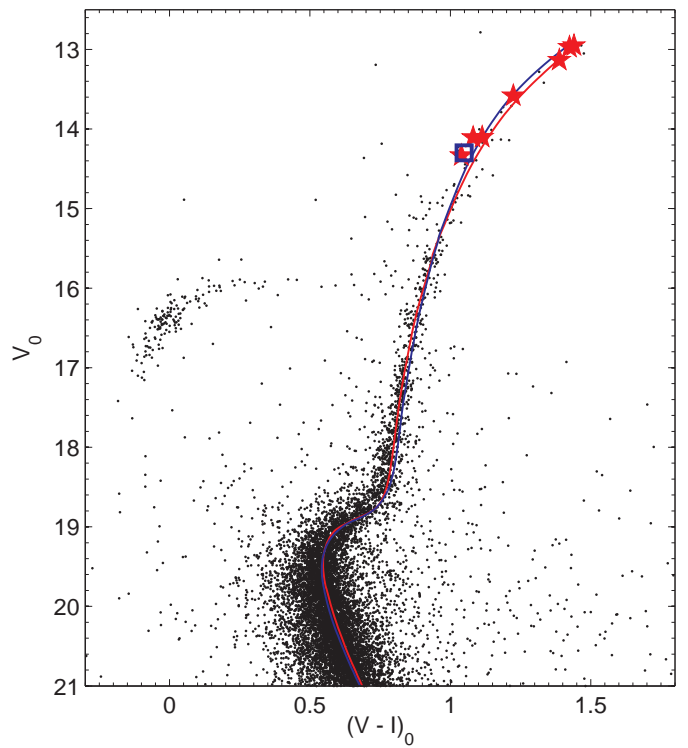}
\caption{CMD of NGC 5897 from Testa et al. (2001); our spectroscopic targets are highlighted as red stars. 
We also indicate the non-member star \#11527 by a blue square. The red line is an $\alpha$-enhanced, 12.4\,Gyr isochrone (Dotter et al. 2008) for [Fe/H]=$-$1.65, 
shifted to a distance modulus of 15.49 and by a reddening vector of E(B$-$V)=0.12, while the blue curve uses our spectroscopic metallicity of [Fe/H]=$-$2.04 and the best distance and reddening
estimates of 15.55 mag and 0.13 mag, respectively.}
\end{figure}

We re-derived the GC's photometric parameters by fitting a grid of isochrones from the Dartmouth database (Dotter et al. 2008) to the cluster's fiducial lines by Testa et al. (2001);
Fig.~1 shows the best-fit isochrone that has an age of 12.4 Gyr, an iron abundance of $-1.65$ dex and [$\alpha$/Fe]=0.4 dex. 
In the following we adopt our  thus derived reddening of E(B$-$V)=0.12
and the reddening laws of Winkler (1997),  as well as the best-fit distance modulus of 15.49 mag. These values are in good agreement with those derived by 
Testa et al. (2001). An independent comparison with isochrones from the Teramo group (Pietrinferni et al. 2004) suggests similar results of 11.5 Gyr, [Fe/H]=$-1.62$, (m$-$M)$_V$=15.41, and 
the same reddening. 
In Figure~1 we also show a metal-poor  isochrone with a metallicity as found from our spectroscopy ([Fe/H]=$-2.04$), with similar distance modulus (15.55 mag), reddening (0.13 mag), and age (13.2 Gyr).
%
%
Where possible, target IDs were taken from Sandage \& Katem (1968) to ease comparison with previous low-resolution studies. 
\subsection{Observations}
Our observations were performed with the Magellan Inamori Kyocera Echelle (MIKE) spectrograph at 
the 6.5-m Magellan2/Clay Telescope at Las Campanas Observatory, Chile.
The data were gathered over four nights in May 2013. By using a slit width of 0.5$\arcsec$ and binning the 
CCD pixels by 2$\times$1, we obtained a resolving power of R$\sim$40,000. 
Here, we used the red and blue arms of the instrument, which  cover a full wavelength range of 3340--9150\AA, 
although we  primarily focused on the red wavelength region above $\sim$4900\AA\ in our analysis. 
Each star was typically exposed for 0.5--2.5 hours, which we split into several exposures to facilitate cosmic ray removal. 
The median seeing was 0.65$\arcsec$ with individual exposures reaching as high as 1$\arcsec$, but notably better conditions ($\sim0.4\arcsec$) during the first night. 
{The seeing variations also prompted the adaptive exposure times for stars of similar magnitudes, as seen in Table~1; 
e.g., for stars S-255 and S-20 the conditions 
changed from 0.85$\arcsec$ to 0.45$\arcsec$.}

The data were processed within the pipeline reduction package of Kelson (2000; 2003), which comprises  
flat field division, order tracing from quartz lamp flats, and 
wavelength calibration using built-in Th-Ar lamp exposures that were taken immediately after each science exposure. 
Continuum-normalisation was obtained by dividing the extracted spectra by a high-order polynomial.  
The resulting spectra reach signal-to-noise (S/N) ratios of 90--120 per pixel as measured from the peak of the order containing H$\alpha$, declining towards $\sim$65 (35) 
in the bluer orders, at 4500\AA~(4000\AA). 
An observing log is given in Table~1, which also lists the basic, photometric properties of the target stars. 

\begin{table*}[htb]
\caption{Properties of the targeted stars.}             
\centering          
\begin{tabular}{rrccccccccc}     
\hline\hline       
& & $\alpha$  & $\delta$ & V & V$-$I & V$-$K & & t$_{\rm exp}$ & S/N\tablefootmark{c} & v$_{\rm HC}$  \\
\raisebox{1.5ex}[-1.5ex]{ID\tablefootmark{a}} &\raisebox{1.5ex}[-1.5ex]{Alt. ID\tablefootmark{b}}& (J2000.0) & (J2000.0)  & [mag] & [mag] & [mag] 
& \raisebox{1.5ex}[-1.5ex]{Date of obs.} & [s] & [pixel$^{-1}$] & [km\,s$^{-1}$]\\
\hline
S-255 &  ESO-1702 & 15:17:30.07  & $-$21:02:08.81 & 13.32 & 1.59 & 3.64 & May 10, 2013  &  6340 & 115  & 104.6 \\ 
S-20$\rlap{\tablefootmark{d}}$ &  ESO-32 & 15:17:14.73  & $-$20:56:57.18 & 13.34 & 1.58 & 3.59 & May 07, 2013      & 1800 & 110 & 100.1\\   
S-9   & ESO-3884 & 15:17:30.51  & $-$20:56:58.90 & 13.51 & 1.54 & 3.51 & May 07, 2013      & 2160 & 120 &100.9 \\  
S-366 & ESO-6910 & 15:17:14.06  & $-$21:04:19.72 & 13.95 & 1.38 & 3.10 & May 07, 2013      & 3600 & 105 & 101.8 \\  
S-138 & ESO-4959 & 15:17:39.38  & $-$20:59:53.25 & 14.48 & 1.27 & 3.06 & May 07, 08, 2013  & 8100 & 120 &101.1  \\  
S-290 & ESO-5660 & 15:17:09.41  & $-$21:01:46.98 & 14.48 & 1.23 & 2.89 & May 10, 2013      & 6600 & 100 & 102.2 \\ 
S-364 &  ESO-1974 & 15:17:21.49  & $-$21:04:40.91 & 14.71 & 1.19 & 2.83 & May 11, 2013      & 5400 & 90 & 103.3 \\ 
\hline
\dots & ESO-4678$\rlap{\tablefootmark{e}}$ & 15:17:45.47  & $-$21:01:54.93 & 14.67 & 1.20 & 2.73 &  May 10, 11 2013 & 5400 & 100& $-$52.7  \\
\hline
\hline                  
\end{tabular}
\tablefoot{
\tablefoottext{a}{Identifications from Sandage \& Katem (1968).}
\tablefoottext{b}{ID from Testa et al. (2001).}
\tablefoottext{c}{Measured in the order containing H$\alpha$.}
\tablefoottext{d}{Variable star$\equiv$V-5 (Sandage \& Katem 1968).}
\tablefoottext{e}{Non-member.}}
\end{table*}
\subsection{Radial velocities}
We measured radial velocities of the targets by cross-correlating the spectra between 5500--6500\AA~against a synthetic spectrum of a red giant 
with stellar parameters representative of the upper RGB of NGC 5897, as expected for our targets. 
This yielded typical uncertainties of $\sim$0.2 km\,s$^{-1}$. 
Five of our targets were included in the low-resolution sample of Geisler et al. (1995), who measured radial velocities and 
metallicities from the near-infrared calcium triplet (CaT). 
On average, we find an excellent agreement with values of Geisler et al. (1995); individual differences of up to 2 km\,s$^{-1}$ could indicate possible 
 binarity of some targets, but since our prime motive is not a detailed kinematic study of this GC,  we did not investigate the velocities  any further. 
Likewise, for the two stars we have in common with the CaT sample of  Rutledge et al. (1997) we measured  velocities larger by $\sim$4 km\,s$^{-1}$. 

One of our targets, ESO-4678, turned out to be a non-member with a highly deviant radial velocity and different metallicity. While listed in Table~1 for completeness, we ignore this object 
for the remainder of this work. All other stars are confirmed cluster members. 
\section{Abundance analysis}
We derived chemical element abundances through a standard, absolute analysis using equivalent widths (EWs),  that  closely follows  the procedures used in our previous works
(e.g., Koch et al. 2009; Kacharov et al. 2013). 
All analyses employed  the 2010 version of the stellar abundance code MOOG (Sneden 1973).  
\subsection{Line list}
The line list for this work builds on that used in Koch \& C\^ot\'e (2010), which, in turn, was assembled from 
various sources (see references therein). Additional transitions for several heavy elements (Zr, La, Ce, Dy) were 
extracted from Shetrone et al. (2003), Sadakane et al. (2004),  and Yong et al. (2005).  

The EWs were measured by fitting a Gaussian profile to the absorption lines 
using IRAF's {\em splot}. 
The final line list for this absolute analysis is provided  in  Table~2. 
Individual elements and transitions will be further discussed in Sect.~5.  
\begin{table*}[hbt]
\caption{Linelist.}             
\centering          
\begin{tabular}{cccrrrrrrrr}     
\hline\hline       
& $\lambda$ & E.P.& & \multicolumn{7}{c}{EW [m\AA]} \\
\cline{5-11}
\raisebox{1.5ex}[-1.5ex]{Element} & [\AA] & [eV]  &\raisebox{1.5ex}[-1.5ex]{log\,$gf$} 
& S-255 & S-20 & S-9 & S-366 & S-138 & S-290 & S-364 \\
\hline                    
$[$O I$]$ & 5577.34 &  1.97 &  $-$8.204 &	5 & \dots & \dots & \dots & \dots & \dots & \dots \\
$[$O I$]$ & 6300.31 &  0.00 &  $-$9.819 &    45 &    32 &    40 &	9 &    23 &    21 &    12 \\
$[$O I$]$ & 6363.78 &  0.02 & $-$10.303 &    16 &    19 &    11 &	4 &	7 &	5 &	6 \\
Na I  & 5682.63 &  2.10 &  $-$0.700 &    18 &    48 &    16 &    38 &    12 &    14 &    14 \\
Na I  & 5688.20 &  2.10 &  $-$0.460 &    60 &    73 &    30 &    64 &    11 &    30 &    21 \\
Na I  & 6154.23 &  2.10 &  $-$1.560 &	7 &    11 & \dots &	7 & \dots &	3 & \dots \\
Na I  & 6160.75 &  2.10 &  $-$1.260 &    14 &    24 & \dots &    12 &	5 &	5 &	6 \\
Mg I  & 5528.42 &  4.35 &  $-$0.481 &   167 &   160 &   161 &   135 &   130 &   129 &   120 \\
Mg I  & 5711.09 &  4.33 &  $-$1.660 &    59 &    56 &    61 &    40 &    38 &    35 &    33 \\
Mg I  & 7387.69 &  5.75 &  $-$1.020 &    14 &	9 &    15 &    11 &    18 &    13 &    10 \\
\hline                  
\end{tabular}
\\Table~2 is available in its entirety in electronic form via the CDS.
\end{table*}

The effects of hyperfine structure were considered 
 for lines of the odd-Z elements Mn I, Cu I, Ba II, La II, and Eu II, 
using data from McWilliam et al. (1995), as well as the splitting for Sc II, V I, and Co I, albeit very small (typically $\la0.04$ dex). 
Finally, the abundance results were placed on the Solar, photospheric scale of 
Asplund et al. (2009).
\subsection{Stellar parameters and atmospheres}
As before, we interpolated the model atmospheres from Kurucz's\footnote{\tt http://cfaku5.cfa.harvard.edu/grids.html} 
grid of one-dimensional 72-layer, plane-parallel, line-blanketed models  
without convective overshoot, assuming local thermodynamic equilibrium (LTE) 
for all species. This model grid incorporates
 the $\alpha$-enhanced opacity distribution functions, 
AODFNEW (Castelli \& Kurucz 2003)\footnote{See {\tt http://wwwuser.oat.ts.astro.it/castelli}.}. 
This is a justified assumption, since the majority of the metal poor Galactic halo GCs (and field stars) are enhanced in the $\alpha$-elements 
by $\approx +0.4$ dex. We expect NGC~5897 to also follow this trend, as was confirmed in retrospective by our measurements of an elevated 
[$\alpha$/Fe]  (Sect.~5.3).   
\subsubsection{Spectroscopic temperatures}
We derived spectroscopic temperatures by forcing excitation equilibrium, i.e., by removing the trend in the abundance from 
the Fe~I lines with excitation potential. In this and the following steps we restricted the analysis to moderately strong lines that meet 
 $-5.5\le\log\left({\rm EW}/\lambda\right)\le-4.5$. 
The spectroscopic  T$_{\rm eff}$ is thus typically precise to within $\pm$100 K, based on the range of reasonable slopes. 
{While deriving the GC's abundance distribution upon relying on the spectroscopic values, we noted strong trends
of some abundances with temperature. Particularly strongly affected were \ion{Fe}{I}  and \ion{Fe}{II} and accordingly 
the ensuing abundance ratios [X/Fe] of many elements.  
Thus the Pearson correlation coefficient, $r$, is $-0.76\pm0.15$ for  \ion{Fe}{I}  and $-0.89\pm0.14$ for \ion{Fe}{II}.   
}
\subsubsection{Photometric temperatures}
Next, photometric temperatures were gauged from stellar (V$-$I)  and (V$-$K) colours; to this end, we applied the the temperature-colour calibrations of Ram\'irez \& Mel\'endez (2005)  
to the photometry of Testa et al. (2001), to which we added the  K-band magnitudes  from the 2MASS (Cutri et al. 2003). 
For this, a metallicity of $-1.7$ dex from previous photometric and CaT-studies 
was used as an initial estimate in the calibrations.  
From this, we found that T$_{\rm eff}$(V$-$K) is cooler than T$_{\rm eff}$(V$-$I) by 66 K (1$\sigma$-scatter of 41 K), but note that   
systematic trends in the (V$-$K)-temperature scale have been reported before (Fabbian et al. 2009; Kacharov et al. 2013).
Overall, our mean photometric values are larger than the  spectroscopic temperatures by 128$\pm$17 K on average. 
{
However, most importantly, there is no systematic trend in the Fe-abundance from the neutral species with photometric temperature ($r=-0.19\pm0.22$), while 
the trend in the ionized species remains. The exact same behaviour is seen, e.g., in the data of Yong et al. (2003; their Fig.~3) and they argue that a combination of stellar parameter
changes can account for such trends rather than resorting to higher-order effects like NLTE corrections. 

In fact, by using the photometric temperature scale instead of the spectroscopic one, 
as requested by the referee, we alleviate the mean ionization imbalance (Sect. 3.2.4) from [\ion{Fe}{I}/\ion{Fe}{II}]=$-0.34$ 
to $-0.12$ dex.
Thus, we continued the analyses by using the mean of both colour indices as photometric T$_{\rm eff}$ (labeled ``phot.'' below)  in our atmospheres.}
The  atmospheric parameters of the red giant sample are listed in Table~3.
\begin{table}[htb]
\caption{Atmospheric parameters.}             
\centering          
\begin{tabular}{rcccccc}     
\hline\hline       
& \multicolumn{4}{c}{T$_{\rm eff}$ [K]} &  & $\xi$  \\
\cline{2-5}
\raisebox{1.5ex}[-1.5ex]{ID}  & (spec.) & (V$-$K)& (V$-$I) & (phot.) & \raisebox{1.5ex}[-1.5ex]{log\,$g$} & [km\,s$^{-1}$]  \\
\hline                    
 S-255 & 4000 &  4077  & 4174 & 4126 & 0.44 & 2.60 \\ 
  S-20 & 4025 &  4101  & 4185 & 4143 & 0.46 & 2.30 \\ 
  S-9  & 4125 &  4146  & 4218 & 4182 & 0.56 & 2.25 \\ 
S-366  & 4300 &  4400  & 4394 & 4397 & 0.92 & 2.00 \\ 
S-138  & 4370 &  4433  & 4551 & 4492 & 1.19 & 1.90 \\ 
S-290  & 4400 &  4566  & 4596 & 4581 & 1.25 & 2.00 \\ 
S-364  & 4450 &  4607  & 4670 & 4639 & 1.38 & 2.05 \\ 
\hline            
\end{tabular}
\end{table}
\subsubsection{Microturbulence}
We determined microturbulent velocities, $\xi$, from the plot of abundances versus reduced width, $\log\left({\rm EW}/\lambda\right)$, 
of neutral iron lines.  
The range of slopes in this fit indicates typical uncertainties in $\xi$ of maximally 0.2 km\,s$^{-1}$. 
{
We show in Fig.~2 the resulting excitation plot and EW plot and the ensuing, near-flat trends in those plots.}
\begin{figure}[htb]
\centering
\includegraphics[width=1\hsize]{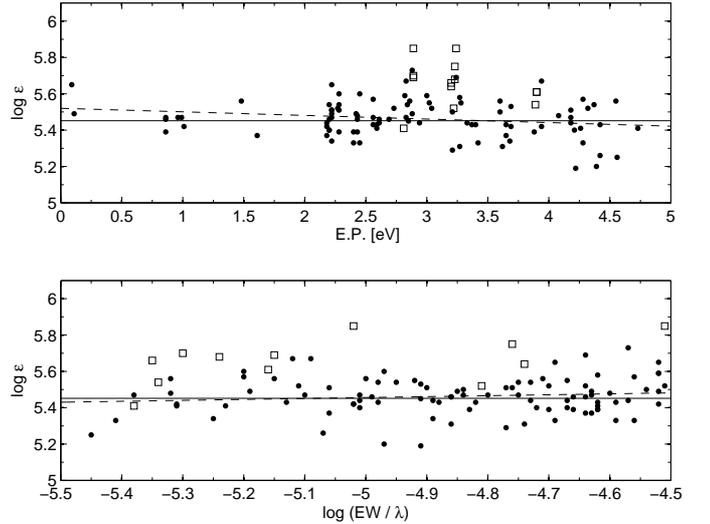}
\caption{Excitation (top panels) and equivalent width (bottom panels) plots for neutral (points) and ionized (open squares) Fe-lines in 
star S-9.
{The solid and dashed lines indicate the mean abundance from the \ion{Fe}{1} lines and the linear regression, respectively.}
}
\end{figure}
\subsubsection{Ionization (non-) equilibrium}
Surface gravities were derived from basic stellar structure considerations (e.g., eq.~1 in  Koch \& McWilliam 2008)  
using {the photometric  T$_{\rm eff}$}, the V-band photometry of Testa et al. (2001), and adopting a distance to NGC~5897 of 12.5 kpc (Harris 2006; Testa et al. 2001), 
which was found to yield a satisfactory fit to the CMD in Fig.~1.
 Comparison of the colors and magnitudes of the target stars 
with the aforementioned isochrones 
indicates an average  stellar mass of 0.79 M$_{\odot}$ for the red giants, which entered in the log\,$g$ determinations. 

Uncertainties in the temperatures ($\pm130$ K), distance ($\pm0.5$ kpc), photometry and reddening  ($\pm0.05$ mag), stellar mass  ($\pm 0.1$ M$_{\odot}$),
 and input metallicity  ($\pm0.1$ dex) imply an average gravity error  of $\sim$0.15 dex.
Ionisation equilibrium is not fulfilled in our stars and we find  a mean difference in 
[Fe\,{\sc i}/Fe\,{\sc ii}] of $-0.12\pm$0.03 dex, while $\varepsilon$(Ti\,{\sc i}) $- \varepsilon$(Ti\,{\sc ii}) is $-0.26\pm0.04$ dex. 
{These values are very similar to those we found in the comparably metal poor GC NGC~6397 (Koch \& McWilliam 2011). 
To test whether the Ti-imbalance could be due to systematic errors in the $gf$-scales, we also employed the most recent laboratory oscillator strengths for
Ti\,{\sc i} (Lawler et al. 2013) and Ti\,{\sc ii} (Wood et al. 2013), which are available for a subset of our line list. These values even deteriorate the non-equilibrium
in Ti\,{\sc i}/{\sc ii} so that we discard this systematic effect as an explanation. In fact, problems with NLTE and ionization equilibrium for Ti have long been suspected in RGB stars 
(e.g., Ruland et al. 1980); cf. Fulbright et al. 2007).
}

We did not attempt to further enforce this equilibrium by adjusting log $g$ (see, e.g.,  Koch \& McWilliam 2008 for an in-depth discussion of the relevant effects).  
In order to mend the discrepancies purely based on gravity, changes would require unrealistically low log $g$ values {as low as $-0.2$ dex for the coolest stars 
and gravities lower by 0.32 dex on average for the entire sample. 
This would settle the iron abundance at equally low values around [Fe/H]$\sim -2$ dex} (see also Sect. 4; Table~4). 
If a false temperature scale was the cause for the disequilibrium, we would need to {decrease the photometric T$_{\rm eff}$
by  120 K  on average, keeping all other parameters constant, to reinstall the ionisation balance.
The imbalance for the warmest stars alone is reconcilable with smaller changes in gravity or temperature that are compatible with the respective errors on those parameters. }
%
%
%
An unrealistically large depletion in the $\alpha$-elements in the atmospheres {of the coolest stars} by $>$1 dex  could yield better agreement of the neutral and ionised species, 
but in the light of the regular [$\alpha$/Fe] ratios we found in the GC stars, we discard this explanation. 

Similarly, we can exclude distance or mass errors as the source of the ionisiation imbalance: 
the stellar mass is well determined from the isochrones and would only have a minor influence on the gravity determinations {if, say, the targets were on the AGB}. 
Likewise, the distance is well constrained from our CMD fit and all measurements agree to within better than 5\%, which is too small to account for the 
required changes in log\,$g$. Furthermore, the reddening from our CMD fit above is consistent with the value derived by Testa et al. (2001);
moreover, the same value was obtained from the maps of 
Schlegel et al. (1998) maps 
with no indication of differential reddening. 

None of the CMD features indicate any need for anomalous He abundances and we consider this an unlikely explanation for the observed imbalance. 
{Thus we conclude in holding with Yong et al. (2003) that a combination of moderate stellar parameter changes is the most likely cause for the 
systematic differences in the Fe\,{\sc i} and Fe\,{\sc ii}  abundances.
}
\subsubsection{NLTE and 3-dimensional corrections}
{
Here we discuss whether the difference between $\varepsilon$(Fe\,{\sc i}) and $\varepsilon$(Fe\,{\sc ii}) could be a consequence of NLTE. 
To this end, we determined abundance corrections for 43 Fe\,{\sc i} and Fe\,{\sc ii} lines from 
 Lind et al. (2012) and  Bergemann et al. (2012). 
As a result, we find that the average neutral iron line abundance is affected at the 0.03 dex-level, while the ionised species suffer downward corrections of only 0.02 dex.
Thus NLTE corrections could alleviate the ionisation non-equilibrium at 0.05 dex. We caution, however, that these calculations were only 
possible for $\xi$=2.0\,km\,s$^{-1}$ and log\,$g$=1.0,  due to the limitations of the respective parameter grid in the sources above so that 
these corrections are slight extrapolations to the parameters of our stars.

Furthermore, we have undertaken an investigation to understand whether unaccounted
3-dimensional (3D) atmosphere effects might explain the ionization imbalance 
 resulting from our 1D analysis.
While work on 3D model stellar atmospheres is not new (e.g., Nordlund 1982;
Nordlund \& Dravins 1990), the work has been focused on the Sun and other dwarf
stars.  However, recent work by Magic et al.  (2013a,b), has extended the 
grid of 3D stellar models to include metal-poor late type stars,
with T$_{\rm eff}$ down to 4000K and metallicities ranging from [Fe/H] $+$0.5
to $-$4.0 dex.  Here, we make use of the grid of mean 3D stellar atmospheres 
(henceforth $<$3D$>$) of Magic et al. (2013b).

Magic et al. (2013a) and Bergemann et al. (2012) compared their $<$3D$>$ results
with those from 1D atmospheres.  Bergemann et al. (2012) used the Magic et al.
(2013b) models to compute NLTE and $<$3D$>$ Fe~I abundance corrections.  The
results indicate significant $<$3D$>$ corrections increasing for hotter, more
metal-poor stars.  In particular, larger corrections were seen for lines of
lower excitation potential.  In 1D LTE analyses, such effects would result in
excitation temperatures, determined from the spectra, significantly higher than
the actual effective temperatures.  Notably, the $<$3D$>$ corrections were smallest
for the Sun and relatively small for HD122563, a cool, metal-poor RGB star
(T$_{\rm eff}$/log\,$g$/[Fe/H] of 4665/1.64/-2.57).

We have employed MOOG to compute the predicted EWs of our Fe~I and Fe~II lines
using two Magic et al. (2013b) $<$3D$>$ models:  with
T$_{\rm eff}$/log\,$g$/[Fe/H]/$\xi$ of 4500 K, 2.00, $-$2.00 dex,
and 1.8 km s$^{-1}$ and 4000 K, 1.50, $-$2.00 dex, and 1.8 km s$^{-1}$.
Unfortunately, the log\,$g$ values of the Magic et al. (2013b) $<$3D$>$ grid only go 
down to 1.50, somewhat higher than our GC RGB stars, whose
gravities range from 0.44 to 1.38.

In our calculations, we employed MOOG and the $<$3D$>$ atmospheres to compute 
predicted EWs for all Fe~I and Fe~II lines measured in our 1D analysis.
 We then followed our 1D abundance spectroscopic analysis method, using Kurucz models,
 to determine T$_{\rm eff}$, $\xi$, and [Fe/H] from the predicted $<$3D$>$ EWs.

For the warmer model, at T$_{\rm eff}$ 4500, log\,$g$ 2.00, it was necessary to
reduce the temperature by 120K in our 1D LTE analysis, to ensure that the
Fe~I abundances were independent of line excitation potential; it was also 
necessary to increase our microturbulent velocity parameter, $\xi$, by 
0.2 km s$^{-1}$, so that the mean iron abundances were independent of EW.
On average, we found that our 1D LTE analysis gave Fe~I abundances 
lower than that input into the $<$3D$>$ EW calculation by 0.09 dex, while the
mean Fe~II abundances were 0.03 dex lower for the 1D model calculations.

For our cooler $<$3D$>$ model, at T$_{\rm eff}$ 4000 and log\,$g$ 1.50, our 1D analysis 
required a temperature lower by 80K, but with the same microturbulent velocity 
as the $<$3D$>$ model, at 1.8 km s$^{-1}$.  In this case, the resulting mean 1D LTE Fe~I abundance was 
lower than the $<$3D$>$ input value by only 0.034 dex, while the mean Fe~II 
abundance was lower by 0.014 dex.  Thus, from our calculations we conclude 
that the $<$3D$>$ contribution to the ionization imbalance for Fe is in the 
range 0.02 to 0.06 dex.

While these $<$3D$>$ results cannot explain the ionization balance in the
majority of our program stars, it is interesting that for the two stars with
parameters closest to our calculations (S-290 and S-364) we measure 1D Fe~I
and Fe~II abundances completely consistent with the small deviations expected
from the $<$3D$>$ models.  Unfortunately, we cannot completely eliminate
$<$3D$>$ effects as the source of our ionization imbalance, because we do not
have $<$3D$>$ models with log\,$g$ near 0.5 to 1.0 at T$_{\rm eff}$ of
4000K and [Fe/H]=$-$2.  Still, our 1D LTE iron ionization imbalance is much 
larger than the computed $<$3D$>$ effect for relatively close stellar parameters;
that is, if $<$3D$>$ effects are responsible, there must be a dramatic increase in the
 effect for the coolest RGB stars. Also, the small dispersion in the Fe~I abundances
 of our sample indicates that the imbalance principally affect the ionized lines
 for the coolest stars.
}
\subsubsection{Model metallicity}
Initially, we assumed a mean metallicity of $-1.7$ dex from the slope of the RGB by Testa et al. (2001) as input for the model atmospheres. 
An independent estimate can be reached from two indicators. 

First, we derived values for [Fe/H]$_{\rm CaT}$ from 
the near-infrared CaT lines at 8498, 8452, and 8662\AA. 
These lines are a well-calibrated metallicity indicator  for red giants in Galactic GCs (Armandroff \& Zinn 1988; 
Rutledge et al. 1997a,b; Carretta \& Gratton 1997; Starkenburg et al. 2010). 
The measurement of the CaT from high-resolution data, however, should be treated with caution:
the dominant abundance information lies in the wings of the lines, while the cores are generally strongly saturated. 
These strong wings are easily affected by errors in the blaze function.
Furthermore, 
any blends with weaker sky residuals, telluric lines, or neutral metal lines will affect their 
shapes; simplistic line profiles usually fail to fit reliably the CaT lines in very high-resolution spectra.  
Since their line cores form in the upper atmosphere and chromosphere, they  are difficult to model reliably (e.g., 
Vernezza et al. 1976; McWilliam et al. 1995; 
 Starkenburg et al. 2010). 
For a simple order-of-magnitude estimate, we numerically integrated the CaT lines 
over the band passes of Armandroff \& Zinn (1988).  
The thus measured EWs were then translated to metallicities on the scale of Carretta \& Gratton (1997) using the 
calibrations of Rutledge et al. (1997a,b) and, as an additional test adopting the calibration of  Starkenburg et al. (2010), which provides a more robust estimate
for metal poor stars.

As a second metallicity indicator, we integrated the Mg I lines at 5167, 5173\AA\ to calibrate [Fe/H] on the  scale of Carretta \& Gratton (1997) as
${\rm [Fe/H]}_{\rm Mg I} =  -2.11 +1.76\,\left[ \Sigma {\rm Mg}   + 0.079\,(V - V_{\rm HB})\right]. $
Here, $\Sigma {\rm Mg}=0.547\,(W_{5167}+W_{5173})$ denotes the Mg index as defined and calibrated in Walker et al. (2007). 
As a result, we find a mean CaT metallicity  [Fe/H]$_{\rm CaT}$ of the seven giants of $-1.84\pm0.01$ dex,  
and  $-1.94\pm0.01$ dex on the scales of  Carretta \& Gratton (1997) and Starkenburg et al. (2010), respectively.  
The latter provides a more reliable calibration at the metal-poor end. 
The mean [Fe/H]$_{\rm Mg~I}$ of our stars is $-1.74\pm0.01$ dex. 
The full range of 0.2 dex in these values illustrates that these methods should only be considered as  order-of magnitude guesses. 
While both the CaT and Mg~I values are listed in Table~7, we point out that these values are only meant as initial metallicity estimates rather than reliable measures of  NGC 5897's abundance scale.

To conclude, 
we used the abundance of the Fe~I lines as input metallicity for the next iteration step as we iterated the above steps simultaneously in all parameters until convergence was reached. 
\subsubsection{Differential analysis}
In order to test further causes of the non-equilibrium, we also performed a differential abundance analysis, relative to a reference star of known stellar parameters.
Here, we chose star \#13414 in NGC 6397 (Koch \& McWilliam 2011), which lies close in parameter space to the present sample star S-9 (T$_{\rm eff}$, log$g$, $\xi$, [Fe/H]  = 
4124 K, 0.29, 1.74, $-2.14$). Differential abundances were then obtained line-by-line using the line list of Koch \& McWilliam (2011) so that uncertainties in stellar parameters, but 
primarily the influence of atomic parameters and potential shortcomings in the model atmospheres  are reduced. 
The list  has $\sim$75, generally weaker, lines in common with the one used in the present study
so that an independent confirmation of our parameters and results can be achieved. As before, we adopt the photometric temperature for this star as granted. 
%


{The main outcome from this exercise is that the microturbulence of S-9 needs to be lowered to 1.53 km\,s$^{-1}$. The overall impact on the abundance results is small in that 
the differential iron abundance of S-9 decreases by $\sim$0.02 dex, while the
difference between neutral and ionized species pertains, at [Fe\,{\sc i}/Fe\,{\sc ii}] = $-0.26$. }
%
%
\subsubsection{The variable star S-20}
Sandage \& Katem (1968) detected variability in star S-20 and classified it  it as a red semiregular variable.
While Wehlau (1990) reports  an overall amplitude of 0.3 mag in V, no period for the variability over the years of 1956--1987 could be measured. 
{
We consulted additional data from the Catalina Surveys (CSS, CRTS; Drake et al. 2009), spanning the years 2006--2013, 
and find regular variations with an amplitude of 0.07 mag in V and  a period of 286 days, 
while individual events are seen as bright as 0.5 mag above average. A detailed analysis of its lightcurve is, however, irrelevant to this work and in the 
following }
 we assume that the stellar parameters of S-20 are sufficiently stable throughout our exposures, which cover a mere 30 minutes,  
and that no systematic changes in the derived abundances were invoked. 
As the results in Sect.~5 imply, this star does not show any striking abundance anomalies or strong deviations from the cluster trends. 
We briefly discuss its abundance patterns within the overall GC environment in Sect.~5.7.
\section{Abundance errors}
To estimate the systematic errors on the chemical  abundance ratios we used the standard approach determined of computing nine new stellar atmospheres 
with altered stellar parameters -- each of  (T$_{\rm eff}$, log\,$g$, $\xi$, [$M$/H], [$\alpha$/Fe]) was varied by its typical uncertainty (Sect.~3.1).  
For testing  the $\alpha$-enhancement, we re-ran the analysis using the solar-scaled opacity distributions, ODFNEW, translating to an uncertainty 
in the models' [$\alpha$/Fe] ratio of 0.4 dex. 
New element ratios were then determined and we list in 
Table~4  the differences in [X/Fe] from those derived using  the best-fit atmospheres as derived in Sect.~3.1. 
This procedure was performed for the stars S-255 and S-364, which bracket the full range in T$_{\rm eff}$. 
\begin{table*}
\caption{Error analysis for the red giants S-255 and S-364.}
\centering          
\begin{tabular}{rccccrccccccrc}
\hline
\hline
& $\Delta$T$_{\rm eff}$ & $\Delta\,\log\,g$ & $\Delta$[M/H] & $\Delta\xi$ & & & & $\Delta$T$_{\rm eff}$ & $\Delta\,\log\,g$ &  $\Delta$[M/H] & $\Delta\xi$ & &  \\
\raisebox{1.5ex}[-1.5ex]{Ion}  & $\pm$100\,K  & $\pm$0.2\,dex  & $\pm$0.1\,dex & $\pm$0.2\,km\,s$^{-1}$ & \raisebox{1.5ex}[-1.5ex]{ODF} & \raisebox{1.5ex}[-1.5ex]{$\sigma_{\rm tot}$} & & 
                                 $\pm$100\,K  & $\pm$0.2\,dex & $\pm$0.1\,dex & $\pm$0.2\,km\,s$^{-1}$ & \raisebox{1.5ex}[-1.5ex]{ODF} & \raisebox{1.5ex}[-1.5ex]{$\sigma_{\rm tot}$} \\
\cline{2-7}\cline{9-14}
& \multicolumn{6}{c}{S-255} & &  \multicolumn{6}{c}{S-364} \\
\hline
\ion{Fe}{I}  & $_{-0.08}^{+0.12}$ & $\mp$0.01  	       & $\mp$0.01	    & $\mp$0.09	         &    0.01 &  0.13 &  &  $\pm$0.15	    & $\mp$0.02  	 & $\mp$0.01	      & $\mp$0.07	   &    0.04 &  0.17\\ 
\ion{Fe}{II} & $_{-0.21}^{+0.11}$ & $_{-0.05}^{+0.03}$ & $\pm$0.02	    & $\mp$0.05	         & $-$0.06 &  0.17 &  &  $_{-0.07}^{+0.05}$ & $_{-0.08}^{+0.06}$ & $\pm$0.02	      & $\mp$0.04	   & $-$0.05 &  0.10 \\ 
\ion{O}{I}   & $\mp$0.01	  & $\pm$0.06	       & $\pm$0.03	    & $<0.01$		 & $-$0.08 &  0.07 &  &  $\pm$0.04          & $\pm$0.08          & $\pm$0.02          & $<0.01$            & $-$0.07 &  0.09 \\ 
\ion{Na}{I}  & $\pm$0.10	  & $_{-0.04}^{+0.03}$ & $\mp$0.02	    & $\mp$0.01 	 &    0.06 &  0.11 &  &  $\pm$0.08          & $\mp$0.02          & $\mp$0.01          & $<0.01$            &    0.03 &  0.08 \\ 
\ion{Mg}{I}  & $_{-0.05}^{+0.08}$ & $\mp$0.03	       & $\mp$0.02	    & $\mp$0.05 	 &    0.04 &  0.09 &  &  $\pm$0.07          & $\mp$0.02          & $\mp$0.01          & $\mp$0.03          &    0.03 &  0.08 \\ 
\ion{Al}{I}  & $\pm$0.09	  & $\mp$0.03	       & $\mp$0.01	    & $<0.01$		 &    0.04 &  0.10 &  &  $\pm$0.08          & $\mp$0.01          & $<0.01$            & $<0.01$            &    0.02 &  0.08 \\ 
\ion{Si}{I}  & $_{-0.08}^{+0.01}$ & $\pm$0.01	       & $\pm$0.01	    & $<0.01$		 & $-$0.02 &  0.05 &  &  $_{-0.01}^{+0.03}$ & $\pm$0.01          & $<0.01$            & $<0.01$            &    0.00 &  0.02 \\ 
\ion{K}{I}   & $_{-0.22}^{+0.19}$ & $_{-0.00}^{+0.02}$ & $\mp$0.01	    & $_{-0.10}^{+0.08}$ & $-$0.00 &  0.23 &  &  $\pm$0.15          & $\mp$0.02          & $\mp$0.02          & $\mp$0.10          &    0.04 &  0.18 \\ 
\ion{Ca}{I}  & $\pm$0.15	  & $\mp$0.03	       & $\mp$0.02	    & $\mp$0.05 	 &    0.05 &  0.17 &  &  $\pm$0.11          & $\mp$0.02          & $\mp$0.01          & $\mp$0.04          &    0.04 &  0.12 \\ 
\ion{Sc}{II} & $_{-0.05}^{+0.01}$ & $\pm$0.04	       & $\pm$0.02	    & $\mp$0.05 	 & $-$0.05 &  0.07 &  &  $\mp$0.01          & $\pm$0.06          & $\pm$0.02          & $\mp$0.03          & $-$0.06 &  0.08 \\ 
\ion{Ti}{I}  & $_{-0.27}^{+0.23}$ & $\mp$0.01	       & $\mp$0.01	    & $\mp$0.09 	 & $-$0.01 &  0.27 &  &  $_{-0.22}^{+0.20}$ & $\mp$0.03          & $\mp$0.02          & $\mp$0.04          &    0.05 &  0.22 \\ 
\ion{Ti}{II} & $_{-0.05}^{+0.02}$ & $\pm$0.01	       & $<0.01$	    & $\mp$0.08 	 & $-$0.03 &  0.09 &  &  $_{-0.02}^{+0.00}$ & $\pm$0.06          & $\pm$0.02          & $_{-0.09}^{+0.06}$ & $-$0.05 &  0.10 \\ 
\ion{V}{I}   & $_{-0.26}^{+0.23}$ & $\mp$0.01	       & $\mp$0.01	    & $\mp$0.01 	 &    0.01 &  0.25 &  &  $_{-0.22}^{+0.20}$ & $\mp$0.03          & $\mp$0.02          & $\mp$0.01          &    0.05 &  0.21 \\ 
\ion{Cr}{I}  & $_{-0.25}^{+0.22}$ & $\mp$0.02	       & $\mp$0.02	    & $_{-0.12}^{+0.10}$ &    0.02 &  0.26 &  &  $\pm$0.19          & $\mp$0.03          & $\mp$0.02          & $\mp$0.07          &    0.06 &  0.20 \\ 
\ion{Mn}{I}  & $_{-0.16}^{+0.18}$ & $\mp$0.01	       & $\mp$0.01	    & $\mp$0.03 	 &    0.02 &  0.17 &  &  $\pm$0.17          & $\mp$0.02          & $\mp$0.01          & $<0.01$            &    0.04 &  0.17 \\ 
\ion{Co}{I}  & $_{-0.09}^{+0.13}$ & $\mp$0.01	       & $<0.01$	    & $\mp$0.02 	 &    0.01 &  0.11 &  &  $\pm$0.16          & $_{-0.02}^{+0.00}$ & $\mp$0.01          & $<0.01$            &    0.04 &  0.16 \\ 
\ion{Ni}{I}  & $_{-0.04}^{+0.09}$ & $\pm$0.01	       & $\pm$0.01	    & $\mp$0.04 	 & $-$0.01 &  0.08 &  &  $\pm$0.12          & $\mp$0.01          & $\mp$0.01          & $\mp$0.02          &    0.03 &  0.13 \\ 
\ion{Cu}{I}  & $_{-0.08}^{+0.12}$ & $\mp$0.01	       & $<0.01$	    & $\mp$0.02 	 &    0.01 &  0.10 &  &  $\pm$0.16          & $\mp$0.02          & $\mp$0.01          & $\mp$0.01          &    0.04 &  0.16 \\ 
\ion{Y}{II}  & $_{-0.02}^{+0.00}$ & $\pm$0.02	       & $\pm$0.01	    & $\mp$0.04 	 & $-$0.03 &  0.05 &  &  $\pm$0.01          & $\pm$0.06          & $\pm$0.02          & $\mp$0.04          & $-$0.05 &  0.07 \\ 
\ion{Ba}{II} & $_{-0.02}^{+0.05}$ & $\pm$0.06	       & $\pm$0.02	    & $_{-0.16}^{+0.13}$ & $-$0.09 &  0.16 &  &  $_{-0.03}^{+0.05}$ & $\pm$0.07          & $\pm$0.02          & $_{-0.16}^{+0.14}$ & $-$0.07 &  0.17 \\ 
\ion{La}{II} & $_{-0.02}^{+0.03}$ & $\pm$0.06	       & $\pm$0.03	    & $\mp$0.01 	 & $-$0.08 &  0.08 &  &  $\pm$0.04          & $\pm$0.07          & $\pm$0.02          & $<0.01$            & $-$0.06 &  0.08 \\ 
\ion{Ce}{II} & $\mp$0.01	  & $\pm$0.05	       & $\pm$0.02	    & $<0.01$		 & $-$0.06 &  0.05 &  &  $\pm$0.03          & $\pm$0.07          & $\pm$0.02          & $<0.01$            & $-$0.05 &  0.08 \\ 
\ion{Pr}{II} & $_{-0.01}^{+0.04}$ & $\pm$0.05	       & $\pm$0.03	    & $\mp$0.01 	 & $-$0.06 &  0.07 &  &  $\pm$0.04          & $\pm$0.07          & $\pm$0.02          & $\mp$0.01          & $-$0.06 &  0.09 \\ 
\ion{Nd}{II} & $\mp$0.02	  & $\pm$0.04	       & $\pm$0.02	    & $\mp$0.02 	 & $-$0.05 &  0.05 &  &  $_{-0.02}^{+0.04}$ & $\pm$0.07          & $\pm$0.03          & $\mp$0.02          & $-$0.05 &  0.08 \\ 
\ion{Sm}{II} & $_{-0.02}^{+0.03}$ & $\pm$0.02	       & $\pm$0.01	    & $\mp$0.02 	 & $-$0.04 &  0.04 &  &  $\pm$0.04          & $\pm$0.06          & $\pm$0.02          & $\mp$0.01          & $-$0.05 &  0.08 \\ 
\ion{Eu}{II} & $_{-0.04}^{+0.00}$ & $\pm$0.07	       & $_{-0.02}^{+0.04}$ & $\mp$0.01 	 & $-$0.08 &  0.08 &  &  $\mp$0.01          & $\pm$0.08          & $\pm$0.03          & $\mp$0.01          & $-$0.07 &  0.08 \\ 
\ion{Gd}{II} & $\mp$0.01	  & $<0.01$	       & $<0.01$	    & $\mp$0.01 	 & $<$0.01 &  0.02 &  &  $\pm$0.03          & $\pm$0.05          & $\pm$0.02          & $\mp$0.01          & $-$0.04 &  0.06 \\ 
\hline                  
\end{tabular}
\end{table*}

The total systematic 1$\sigma$ uncertainty, $\sigma_{\rm tot}$, is obtained by summing all contributions in quadrature\footnote{For this we adopted a quarter of the ``ODF'' error such as to mimic 
an uncertainty of the stars [$\alpha$/Fe] ratio of 0.1 dex.}, although 
 this yields  only a conservative upper limit; the realistic,  underlying errors will be smaller due to the covariances 
of the atmosphere parameters, in particular, between temperature and gravity (e.g., McWilliam et al. 1995).
 As this exercise shows, the total systematic error on [Fe/H] is 0.15 dex from the neutral and of the same order of magnitude for the ionized   
species. 
The $\alpha$-elements are uncertain to within 0.12 dex on average, but we refer to Table~4 to note differences between individual 
elements such as smaller errors for Si compared to larger ones on the  [Ti/Fe] ratios. 
Likewise, the heavier, iron peak and neutron 
capture, elements show systematic errors of 0.07--0.16 dex (quartile range), with the largest values for the iron-peak elements
V and Cr. 
This is in line with the low excitation potential of the lines of these elements, which leads to a strong dependence of their abundances on T$_{\rm eff}$ 
(e.g., Ram\'irez \& Cohen (2003).  
Of all elements, Ba  shows the strongest trend with microturbulence due to its overall strong lines. 
None of our results is strongly affected by changes in the input metallicties, [M/H], 
while changes in the [$\alpha$/Fe] ratio of the atmospheres have a larger impact only on the ionized  species of the heavier elements at Z$\ge$56. 

To account for the random error due to uncertainties in 
the atomic parameters (mainly log\,$gf$), the spectral noise, and potential insufficiencies in the atmosphere models themselves, 
we list in Table~5 the 1$\sigma$ line-to-line scatter and the number of lines used to derive each abundance ratio.
For elements with many measured transitions, like Fe, Ca, Sc, Ti, Cr, or Ni, systematic uncertainties are  the most dominant contribution,
while elements with only a few well-measured lines, the line-to-line scatter will be on par or even dominate the error budget. 
If  only one line was detectable for a certain element, we
conservatively  adopted an uncertainty of 0.10 dex.  For the following plots and statistics, all other elements were assigned a minimum random error of 0.05  dex (e.g., Ram\'irez \& Cohen 2003). 
 \section{Abundance results}
Table~5 lists the abundance ratios relative to Fe I, except for all ionized  species and oxygen, which was measured from the [O I]  lines. These are given relative to Fe II. 
{This table also contains the average abundances and 1$\sigma$ dispersion of the entire RGB sample.}
Our results are also illustrated in the boxplot of Fig.~3 which shows the median and interquartile ranges for each element abundance ratio. 
\begin{figure}[htb]
\centering
\includegraphics[width=1\hsize]{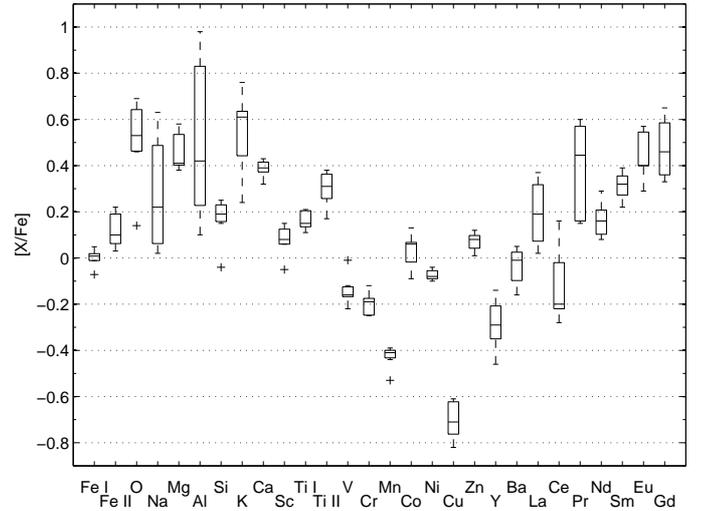}
\caption{Boxplot showing the median {(solid horizontal lines)} and interquartile {(25--75\%)} ranges of the chemical abundance ratios in NGC~5897 {with the full abundance 
coverage depicted by the dashed error-bars}. {Plus symbols designate 1$\sigma$-outliers}.  All values are shown relative to the respective ionization stage, except for 
Fe\,{\sc i} (relative to the cluster mean), Fe\,{\sc ii} (relative to Fe\,{\sc i}), and O\,{\sc i} (relative to Fe\,{\sc ii}.)
}
\end{figure}
\begin{table*}[htb]
\caption{Abundance results {for individual stars and the entire sample}.}             
\centering          
\begin{tabular}{crcrcrcrcrcrcrcr}     
\hline\hline       
& & S-255 & & & & S-20 & & & & S-9 & & & & S-366 &  \\
\cline{2-4}\cline{6-8}\cline{10-12}\cline{14-16}
\raisebox{1.5ex}[-1.5ex]{Element$^a$} & [X/Fe] & $\sigma$ & N &  & [X/Fe] & $\sigma$ & N &  & [X/Fe] & $\sigma$ & N &  & [X/Fe] & $\sigma$ & N \\
\hline                    
Fe$_{\rm CaT}^b$  &$-$1.95 &\dots& \dots &    & $-$1.88 & \dots& \dots &    & $-$1.88 &\dots &\dots &    & $-$1.95 &\dots &  \dots  \\
Fe$_{\rm Mg I}^c$ & $-$1.73 & \dots & \dots &    & $-$1.76 & \dots & \dots &    & $-$1.70 & \dots & \dots &    & $-$1.77 & \dots & \dots     \\
Fe\,{\sc i}       &   $-$1.99 &  0.23 & 133 & &    $-$2.05 &  0.21 & 136 & &	$-$2.05 &  0.14 & 129 & &    $-$2.02 &  0.20 & 131 \\ 
Fe\,{\sc ii}      &   $-$1.77 &  0.11 &  13 & &    $-$1.89 &  0.11 &  12 & &	$-$1.85 &  0.12 &  13 & &    $-$1.92 &  0.20 &  13 \\ 
A(Li)             &  $<-$0.66 & \dots & 1 & &     $<-$0.62 & \dots & 1 &  & $<-$0.50 & \dots & 1 &  & $<-$0.19 & \dots & 1 \\
O\,{\sc i}        &	 0.47 &  0.02 &   3 & &       0.53 &  0.16 &   2 & &	   0.46 &  0.09 &   2 & &	0.14 &  0.07 &   2 \\ 
Na\,{\sc i}\rlap{$^{\rm NLTE}$}&	 0.20 &  0.19 &   4 & &       0.56 &  0.06 &   4 & &	   $-$0.01 &  0.06 &   2 & &	0.47 &  0.09 &   4 \\ 
Mg\,{\sc i}       &	 0.40 &  0.19 &   3 & &       0.41 &  0.22 &   3 & &	   0.56 &  0.19 &   3 & &	0.41 &  0.21 &   3 \\ 
Al\,{\sc i}       &	 0.10 & \dots &   1 & &       0.78 &  0.03 &   4 & &	   0.27 & \dots &   1 & &	0.98 &  0.09 &   4 \\ 
Si\,{\sc i}       &   $-$0.04 &  0.29 &   4 & &       0.25 &  0.26 &   4 & &	   0.24 &  0.29 &   4 & &	0.20 &  0.30 &   4 \\ 
K\,{\sc i}        &	 0.24 &  0.01 &   2 & &       0.39 &  0.03 &   2 & &	   0.61 &  0.06 &   2 & &	0.64 &  0.07 &   2 \\ 
Ca\,{\sc i}       &	 0.32 &  0.18 &  17 & &       0.40 &  0.19 &  17 & &	   0.39 &  0.11 &  17 & &	0.42 &  0.08 &  17 \\ 
Sc\,{\sc ii}      &   $-$0.05 &  0.07 &   7 & &       0.08 &  0.09 &   7 & &	   0.06 &  0.06 &   7 & &	0.15 &  0.09 &   7 \\ 
Ti\,{\sc i}       &	 0.15 &  0.23 &  18 & &       0.11 &  0.22 &  18 & &	   0.13 &  0.20 &  18 & &	0.15 &  0.19 &  18 \\ 
Ti\,{\sc ii}      &	 0.17 &  0.11 &   6 & &       0.25 &  0.14 &   6 & &	   0.37 &  0.23 &   6 & &	0.31 &  0.19 &   6 \\ 
V\,{\sc i}        &   $-$0.16 &  0.16 &   6 & &    $-$0.12 &  0.14 &   6 & &	$-$0.16 &  0.15 &   6 & &    $-$0.17 &  0.10 &   3 \\ 
Cr\,{\sc i}       &   $-$0.25 &  0.09 &   7 & &    $-$0.19 &  0.11 &   7 & &	$-$0.12 &  0.13 &   7 & &    $-$0.17 &  0.08 &   7 \\ 
Mn\,{\sc i}       &   $-$0.39 &  0.13 &   6 & &    $-$0.41 &  0.15 &   6 & &	$-$0.41 &  0.10 &   6 & &    $-$0.40 &  0.05 &   6 \\ 
Co\,{\sc i}       &	 0.06 &  0.11 &   4 & &       0.13 &  0.10 &   4 & &	   0.06 &  0.13 &   4 & &    $-$0.01 &  0.14 &   4 \\ 
Ni\,{\sc i}       &   $-$0.09 &  0.22 &  26 & &    $-$0.05 &  0.24 &  26 & &	$-$0.04 &  0.24 &  26 & &    $-$0.08 &  0.22 &  26 \\ 
Cu\,{\sc i}       &   $-$0.71 &  0.13 &   3 & &    $-$0.61 &  0.05 &   3 & &	$-$0.71 &  0.11 &   2 & &    $-$0.62 &  0.04 &   2 \\ 
Zn\,{\sc i}       &	 0.09 &  0.11 &   2 & &       0.10 &  0.06 &   2 & &	   0.12 &  0.19 &   2 & &	0.08 &  0.08 &   2 \\ 
Y\,{\sc ii} 	  &   $-$0.46 &  0.20 &   5 & &    $-$0.29 &  0.14 &   5 & &	$-$0.35 &  0.13 &   5 & &    $-$0.19 &  0.10 &   5 \\ 
Ba\,{\sc ii}	  &   $-$0.16 &  0.03 &   3 & &    $-$0.12 &  0.02 &   3 & &	$-$0.03 &  0.11 &   3 & &	0.01 &  0.10 &   3 \\ 
La\,{\sc ii}	  &	 0.02 &  0.13 &   5 & &       0.19 &  0.08 &   5 & &	   0.11 &  0.09 &   5 & &	0.28 &  0.08 &   5 \\ 
Ce\,{\sc ii}	  &   $-$0.22 & \dots &   1 & &    $-$0.28 & \dots &   1 & &	$-$0.22 & \dots &   1 & &    $-$0.20 & \dots &   1 \\ 
Pr\,{\sc ii}	  &	 0.15 & \dots &   1 & &       0.43 & \dots &   1 & &	   0.16 & \dots &   1 & &	0.57 & \dots &   1 \\ 
Nd\,{\sc ii}	  &	 0.08 & \dots &   1 & &       0.22 & \dots &   1 & &	   0.09 & \dots &   1 & &	0.16 & \dots &   1 \\ 
Sm\,{\sc ii}	  &	 0.22 &  0.07 &   6 & &       0.36 &  0.07 &   5 & &	   0.26 &  0.05 &   5 & &	0.34 &  0.08 &   6 \\ 
Eu\,{\sc ii}	  &	 0.40 &  0.05 &   2 & &       0.40 &  0.02 &   2 & &	   0.40 &  0.07 &   2 & &	0.56 &  0.06 &   2 \\ 
Gd\,{\sc ii}      &	 0.45 & \dots &   1 & &       0.33 & \dots &   1 & &	   0.33 & \dots &   1 & &	0.46 & \dots &   1 \\ 
\hline                  
& &  S-138 & & & & S-290 & & & & S-364 & & & \multicolumn{2}{c}{NGC 5897} \\
\cline{2-4}\cline{6-8}\cline{10-12}\cline{14-15} 
\raisebox{1.5ex}[-1.5ex]{Element$^a$} & [X/Fe] & $\sigma$ & N &  & [X/Fe] & $\sigma$ & N &  & [X/Fe] & $\sigma$ & N & & $\langle$\,[X/Fe]\,$\rangle$ & $\sigma$ & \\
\hline                    
Fe$_{\rm CaT}^b$  & $-$1.97 & \dots & \dots &  & $-$2.00 & \dots & \dots & &  $-$1.97 & \dots & \dots &	  & $-$1.94 &  0.05  & \\
Fe$_{\rm Mg I}^c$ & $-$1.74 & \dots & \dots &  & $-$1.73 & \dots & \dots & &  $-$1.71 & \dots & \dots &   & $-$1.73 &  0.03  & \\
Fe\,{\sc i}     &   $-$2.11 &  0.11 & 121 & &    $-$2.02 &  0.16 & 136 & &    $-$2.03 &  0.15 & 129 & &     $-$2.04 &  0.04  & \\ 
Fe\,{\sc ii}    &   $-$2.01 &  0.12 &  12 & &    $-$1.99 &  0.12 &  13 & &    $-$1.98 &  0.12 &  13 & &     $-$1.92 &  0.09  & \\
A(Li)           &  $<-$0.23 & \dots &   1 & &   $<-$0.08 & \dots &   1 & &   $<-$0.15 & \dots &   1 & &       \dots & \dots  & \\
O\,{\sc i}      &      0.69 &  0.04 &   2 & &       0.65 &  0.15 &   2 & &       0.62 &  0.12 &   2 & &        0.51 &  0.19  & \\
Na\,{\sc i}\rlap{$^{\rm NLTE}$}& $-$0.03 & 0.20 & 3 & &   0.14 &  0.06 &   4 & &       0.16 &  0.11 &   3 & &        0.27 &  0.24  & \\
Mg\,{\sc i}     &      0.58 &  0.30 &   3 & &       0.46 &  0.23 &   3 & &       0.38 &  0.17 &   3 & &        0.46 &  0.08  & \\
Al\,{\sc i}     &   \dots &  \dots &\dots & &    \dots &  \dots &\dots & &       0.42 &  0.15 &   2 & &        0.51 &  0.36  & \\
Si\,{\sc i}     &      0.19 &  0.22 &   4 & &       0.18 &  0.34 &   4 & &       0.15 &  0.31 &   4 & &        0.17 &  0.10  & \\
K\,{\sc i}      &      0.76 &  0.23 &   2 & &       0.60 &  0.17 &   2 & &       0.62 &  0.12 &   2 & &        0.55 &  0.18  & \\
Ca\,{\sc i}     &      0.43 &  0.12 &  17 & &       0.38 &  0.10 &  17 & &       0.37 &  0.13 &  17 & &        0.39 &  0.04  & \\
Sc\,{\sc ii}    &      0.13 &  0.08 &   7 & &       0.11 &  0.08 &   7 & &       0.06 &  0.11 &   7 & &        0.08 &  0.07  & \\
Ti\,{\sc i}     &      0.21 &  0.21 &  17 & &       0.19 &  0.17 &  17 & &       0.21 &  0.15 &  17 & &        0.16 &  0.04  & \\
Ti\,{\sc ii}    &      0.34 &  0.15 &   6 & &       0.38 &  0.16 &   6 & &       0.28 &  0.14 &   6 & &        0.30 &  0.07  & \\
V\,{\sc i}      &   $-$0.22 &  0.06 &   2 & &    $-$0.01 &  0.04 &   2 & &    $-$0.14 & \dots &   1 & &     $-$0.14 &  0.07  & \\
Cr\,{\sc i}     &   $-$0.25 &  0.04 &   7 & &    $-$0.19 &  0.05 &   7 & &    $-$0.24 &  0.12 &   7 & &     $-$0.20 &  0.05  & \\
Mn\,{\sc i}     &   $-$0.53 &  0.10 &   5 & &    $-$0.44 &  0.14 &   5 & &    $-$0.40 &  0.07 &   5 & &     $-$0.43 &  0.05  & \\
Co\,{\sc i}     &      0.07 &  0.09 &   4 & &    $-$0.02 &  0.09 &   4 & &    $-$0.09 &  0.09 &   3 & &        0.03 &  0.07  & \\
Ni\,{\sc i}     &   $-$0.07 &  0.23 &  25 & &    $-$0.09 &  0.23 &  24 & &    $-$0.10 &  0.25 &  25 & &     $-$0.07 &  0.02  & \\
Cu\,{\sc i}     &   $-$0.82 &  0.12 &   2 & &    $-$0.78 & \dots &   1 & &    $-$0.63 & \dots &   1 & &     $-$0.70 &  0.08  & \\
Zn\,{\sc i}     &      0.08 &  0.03 &   2 & &       0.01 &  0.09 &   2 & &       0.03 &  0.04 &   2 & &        0.07 &  0.04  & \\
Y\,{\sc ii} 	&   $-$0.35 &  0.11 &   5 & &    $-$0.14 &  0.12 &   5 & &    $-$0.26 &  0.14 &   5 & &     $-$0.29 &  0.11  & \\
Ba\,{\sc ii}	&      0.05 &  0.17 &   3 & &       0.03 &  0.08 &   3 & &    $-$0.01 &  0.04 &   3 & &     $-$0.03 &  0.08  & \\
La\,{\sc ii}	&      0.06 &  0.19 &   4 & &       0.37 &  0.18 &   4 & &       0.33 &  0.13 &   4 & &        0.19 &  0.14  & \\
Ce\,{\sc ii}	&   $-$0.02 & \dots &   1 & &    $-$0.02 & \dots &   1 & &       0.16 & \dots &   1 & &     $-$0.11 &  0.16  & \\
Pr\,{\sc ii}	&      0.60 & \dots &   1 & &    \dots &  \dots &\dots & &       0.46 & \dots &   1 & &        0.39 &  0.20  & \\
Nd\,{\sc ii}	&      0.14 & \dots &   1 & &       0.17 & \dots &   1 & &       0.29 & \dots &   1 & &        0.16 &  0.07  & \\
Sm\,{\sc ii}	&      0.31 &  0.11 &   4 & &       0.32 &  0.10 &   5 & &       0.39 &  0.05 &   6 & &        0.31 &  0.06  & \\
Eu\,{\sc ii}	&      0.29 &  0.01 &   2 & &       0.57 &  0.03 &   2 & &       0.50 & \dots &   1 & &        0.45 &  0.10  & \\
Gd\,{\sc ii}	&      0.51 & \dots &   1 & &       0.65 & \dots &   1 & &       0.61 & \dots &   1 & &        0.48 &  0.12  & \\

\hline                  
\end{tabular}
\begin{flushleft}
$^a$Ionized  species and O are given relative to Fe\,{\sc ii}. Abundance ratios are listed relative to iron, safe for Fe\,{\sc i}  and Fe\,{\sc ii} (relative to H) and 
 Li. 
\\$^b$Metallicity estimate based on the calcium triplet calibration of Starkenburg et al. (2010). 
\\$^c$Metallicity estimate based on the Mg I calibration of Walker et al. (2007), on the metallicity scale of Carretta \& Gratton (1997).
\end{flushleft}
\end{table*}
\subsection{Iron}
Our study resulted in a mean iron abundance from the neutral species of [Fe/H]=$-2.04\pm0.01$ (stat.) $\pm0.15$ (sys.), and
 [Fe/H]=$-1.92\pm0.03$ (stat.) $\pm0.15$ (sys.) from the ionized lines.
All previous studies have reported higher values for the GC's metallicity: 
Testa et al. (2001) obtained  [Fe/H] = $-1.7\pm0.16$ from the slope of RGB; 
similar values are found on the scales of by Zinn \& West (1984) and Rutledge et al. (1997), viz. $-1.68$ and $-1.73$ dex, respectively. 
The most comprehensive  low-resolution spectroscopic study of this GC to date is that of Geisler et al. (1995), who measured 
a mean metallicity of  $-1.94\pm0.04$  from the CaT in 14 red giants. 
More recent spectroscopic data appear to converge on a more metal poor mean compared to earlier photometric studies and we note here 
the value by Carretta et al. (2009a) of $-1.90\pm0.06$ and the scale of Kraft \& Ivans (2003) that reaches as low as $-2.09$ for NGC~5897.

Also Sobeck et al. (2011) report on a downward correction of the [Fe/H] value by $\sim$0.3 dex of the metal poor GC M15 compared to older spectroscopic studies. 
Furthermore, they found a discrepancy of up to 0.15 dex between the metallicities of their red HB and their RGB sample;  reasons for the difference were sought 
in gravitational settling or the missing treatment of sphericity in the atmospheres, while uncertainties in the 
atmospheric parameters or atomic physics were excluded as possible causes. While we cannot test any systematic differences with evolutionary status in NGC~5897 with our 
pure RGB sample, downward corrections at the metal poor end do not appear uncommon.

None of the studies mentioned above leaves any room for a significant iron- or metallicity  spread and this is not found in our data, neither. NGC~5897's 
intrinsic 1$\sigma$-dispersion of  $0.04\pm0.01$ dex
is entirely explicable with the mass of this GC ($M_V=-7.23$), which allows for a minimum amount of stochastic, internal enrichment in its central potential well
(e.g., Carretta et al. 2009a). 
\subsection{Lithium}
We determined upper limits on A(Li) from the region around the resonance line at 6707 \AA, using the noise characteristics and expected extent of the absorption feature, which 
is usually not visible in spectra of cool, red giants.  
Our  low 3$\sigma$ upper limits of A(Li)$\la0.15$ are thus compatible with the full destruction of Li in the stellar interiors, as is also found in 
red giants of similar effective temperatures and luminosities in other GCs over a broad range of metallicities  
(e.g., Lind et al. 2009). 
\subsection{Light elements: O, Na, Al, K}
{The [O I] 6300\AA-line can potentially suffer from contamination with telluric lines. To test this, we compared our spectra 
with spectra of the telluric standard star HR~7889, taken under the same conditions as the GC targets. Thus we confirm that 
the immediate region around the line in our stars is not affected by any telluric contamination.}

{NLTE corrections for sodium are scarcely available in the literature for our exact stellar parameter combinations, in particular the low gravities. 
Using the data by Lind et al. (2011)\footnote{Taken from the authors' web-based database, {\tt www.inspect-stars.net}.} for the warmer stars we confirmed that the corrections are of the 
order of $-0.05$ dex, averaged over the, usually four, Na-lines we used in the analysis.}
Figure~4 illustrates the well-sampled Na-O anti-correlation in NGC~5897, in comparison with the comprehensive sample for Galactic GCs compiled by 
Carretta et al. (2009b,c). Thus NGC~5897 fully conforms with the bulk of GCs in that it shows chemical evidence for multiple stellar populations. 
\begin{figure}[htb]
\centering
\includegraphics[width=1\hsize]{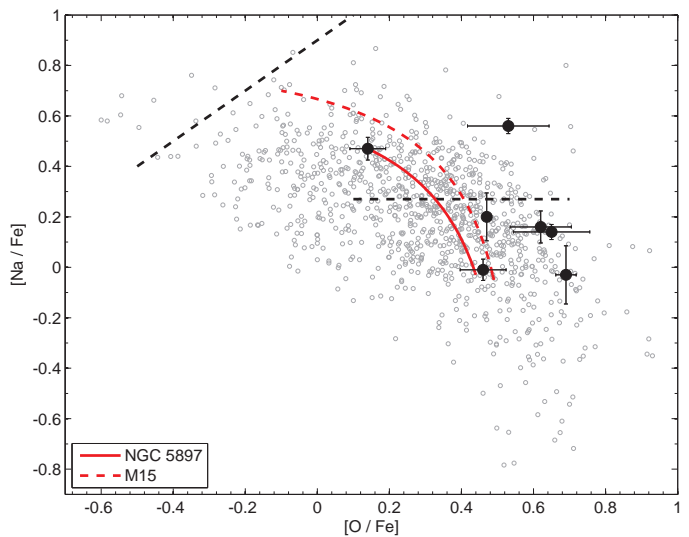}
\caption{Na-O anti-correlation in NGC~5897 (solid circles) on the data for Galactic GCs of Carretta et al. (2009b,c;  gray points). {The Na-abundances have been corrected for NLTE, with typical corrections of the order of $-0.05$ dex.} The dashed lines are empirical separations into primordial (lower part), intermediate (middle), and extreme (upper left corner) 
populations. Shown as as solid red line is the best-fit simple dilution model for NGC 5897, compared to the curve of the metal poor GC M~15 from Carretta et al. (2009c; dashed red line).}
\end{figure}
Here, we indicate the empirical separations into the different generations of stars, established by Carretta et al. (2009c). 
The majority of our sample can be dubbed primordial (P), as they are falling below [Na/Fe]$_{\rm min}$+0.3 dex, thus exhibiting the low Na- and 
large O-enhancements typical of stars that was only affected by the ejecta from early, massive SNe II without any contributions to the Na-budget. 
These coincide with the parameter space spanned by Galactic halo field stars.  
The remaining two giants already show indications of p-capture reactions that took place in the first-generation stars that ultimately enriched the ISM to form 
the second generation. Accordingly, these two giants of the Intermediate (I) population are enhanced in Na, with signs of lowered O-levels. 
We note that the variable S-20 is characterized by the largest Na/Fe ratio of our sample, while its [O/Fe] ratio is not significantly depleted 
and also overlaps with the P-generation. 

Several studies have indicated that the ratio of first-to-second generation stars is approximately 30:70 (e.g., Carretta 2013).
The fraction  of 5:2 in our sample would rather imply the opposite trend, i.e., 70\% of the stars belonging to the first generation.  
We caution, however, that this argument is hampered by our relatively sparse sample. The distinction into P and I components based on our seven giants 
may be inadequate as we may be 
missing first-generation stars even more depleted in  Na. Thus lowering [Na/Fe]$_{\rm min}$ would actually assign more of our targets to the second generation, 
thereby boosting the ratio to be more in line with the majority of Galactic GCs studied to date. 

The same holds when assessing the enrichment processes in this GC by computing simple dilution models (Carretta et al. 2009c). While we obtain a best fit to our few stars 
for [Na/Fe]$_{\rm max}=4.57$ and  [O/Fe]$_{\rm max}=0.44$, a broader, true range in the Na-O abundance space cannot be excluded at present. 
Interestingly, NGC~5897 bears close resemblance to the Galactic halo GC M15, not only in terms of its metal poor nature ([Fe/H]$_{\rm M15} = -2.6$) and its location within the halo 
(R$_{\rm GC}$=10.4 kpc), but also when considering the dilution models in  Na-O space in Fig.~4. 
Other, similarly metal-poor GCs covered by Carretta et al. (2009c)  show much a broader range in the extrema of Na and/or O,  and/or steeper slopes in the Na-O anti-correlation, such as NGC~6397. 
This suggests that a variety of factors, besides metallicity,  is at play in shaping the details of the stellar generations seen in the GCs.

We note that Al measurements were only possible for five of the giants and that no clear Mg-Al correlation is discernible. 
A common trend is maintained in that the star with the lowest O-abundance  also shows the largest enhancement in Al. 
{
 There is other clear proof of the action of p-capture reactions among stars 
 of the sample in NGC 5897:
 there is a clear Na-Al correlation  (bottom panel of Fig.~5) in that  the two giants with the highest Na abundance also have the highest Al
 abundance. 
This  additional evidence strongly supports the presence of  multiple stellar generations in this cluster.
We note that any trend in Mg vs. O (top panel of Fig.~5) would be mainly driven by the O-poor star S-366, which, however,  still has a [Mg/Fe] ratio
compatible with the bulk of the GC stars.
A correlation could be expected,  since both elements
 are depleted in p-capture reactions. 
}
\begin{figure}[htb]
\centering
\includegraphics[width=1\hsize]{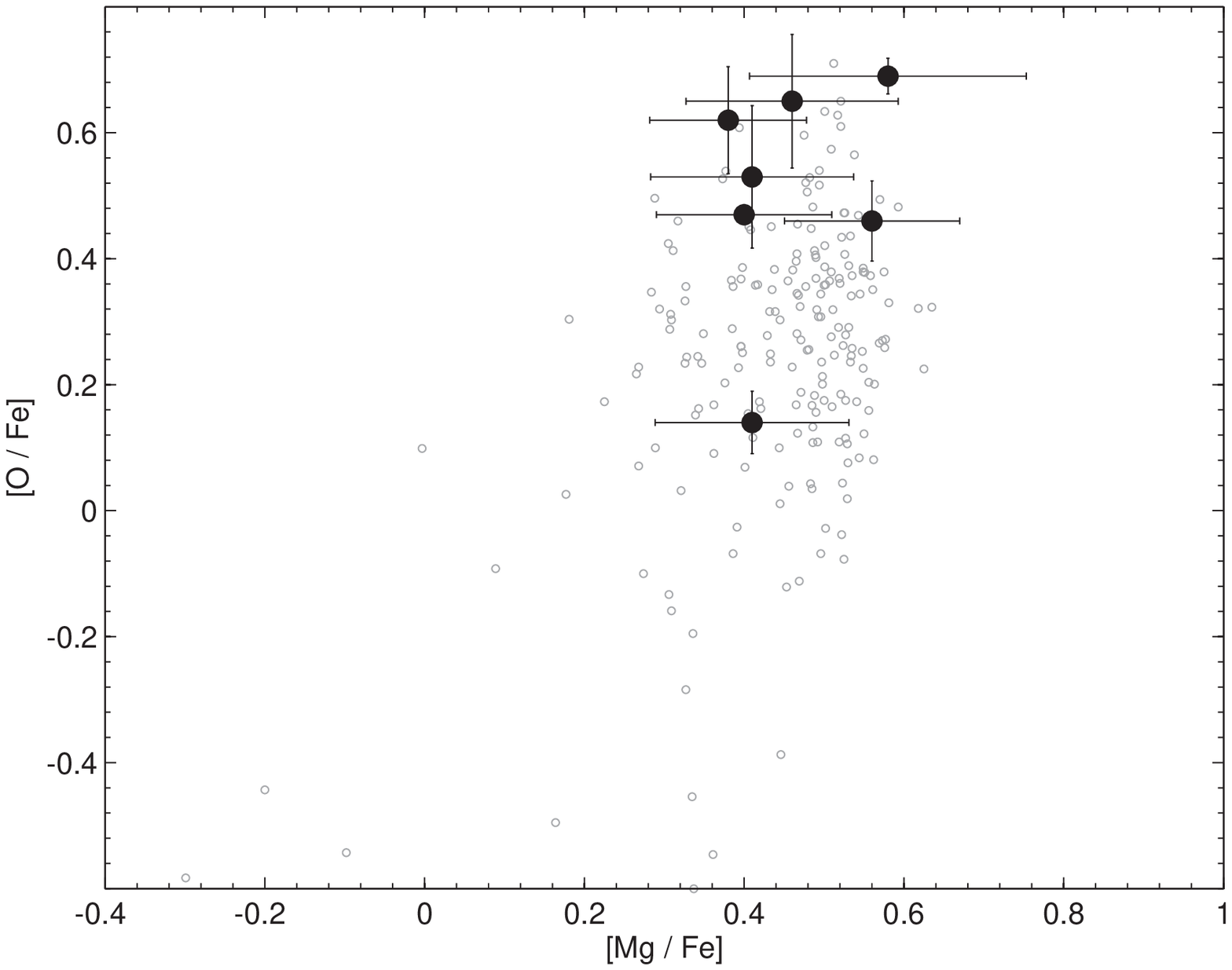}
\includegraphics[width=1\hsize]{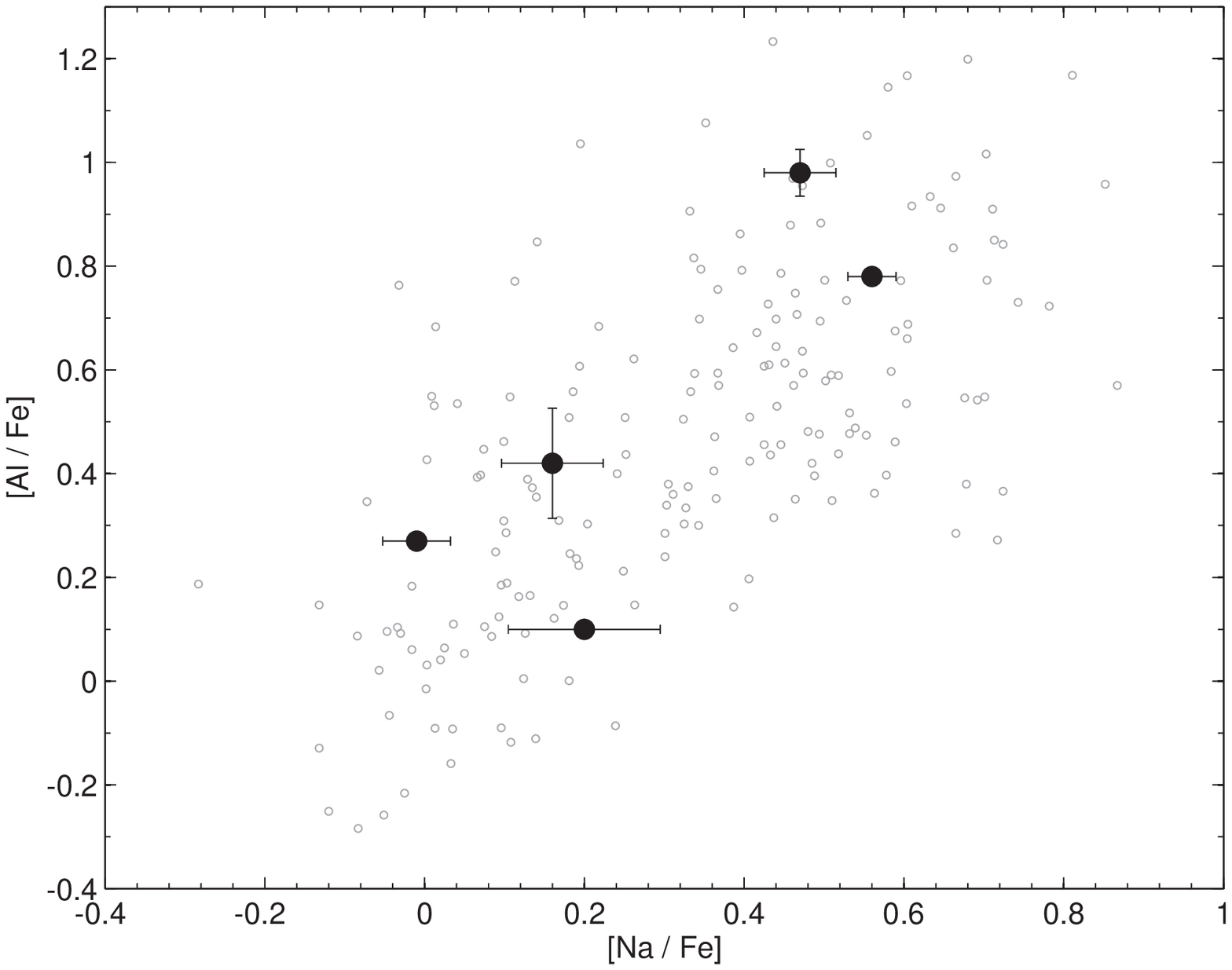}
\caption{Other p-capture element correlations using the same data as in Fig.~4.}
\end{figure}

Finally, no firm conclusions on trends of the potassium abundances can be reached from the strong resonance lines at 7664, 7698\AA, as we do not sample 
the full extent of this GC's stellar populations (e.g., Mucciarelli et al. 2012) and those lines generally require a more detailed NLTE treatment (e.g., Takeda et al. 2009). 
\subsection{$\alpha$-elements: Mg, Si, Ca, Ti}
With regard to the elements primarily produced in type II supernovae (SNe), NGC~5897 bears little surprises.
We find a mean [$\alpha$/Fe] = $<[$Mg,Ca,Ti I / 3 Fe I$]> = 0.34\pm0.01$(stat.)$\pm0.12$(sys.), 
fully in line with the $\alpha$-enhancement found in Galactic halo GCs and field stars (e.g., Venn et al. 2004; Pritzl et al. 2005), 
rendering this object fully representative of its metallicity (e.g., Koch \& McWilliam 2011).   
NGC 5897 is also not unusual in that the 
intrinsic scatter in the [$\alpha$/Fe] ratios is below well 0.1 dex and thus explicable with the measurement errors.

Due to their production in the  explosive SNe II events (Ca, Si, Ti) and the hydrostatic burning of the progenitors (Mg, O), the $\alpha$-element
abundances can be expected to trace each other, as is indeed found in all major Galactic components (e.g., Fulbright et al. 2007). 
In this context, we measured [Mg/Ca] and [Ca/Si] ratios that are slightly elevated by 0.07 dex (1$\sigma$ scatter of 0.07 dex), 
which compares to the values of around zero in Galactic disk and halo stars. 
However, the [Ca/Ti] ratio in our stars is notably higher, at 0.22 dex, and with a 1$\sigma$ scatter of 
0.05 dex also remarkably homogeneous. As was argued, e.g.,  in Koch \& McWilliam (2010) this can be caused by an  NLTE overionization of Ti I. 

Star S-255 has $\alpha$/Fe ratios lower than the GC average, notably in Si and, to a lesser extent, in Ca, 
but it does not stand out in the relative ratios of different $\alpha$-elements.
\subsection{Iron-peak- and heavy elements: \hspace{4cm}Sc, V, Mn, Cr, Co, Ni, Cu, Zn}
As for the $\alpha$-elements, none of the Fe-peak element distributions in NGC~5897 is outstanding, neither.
Due to the same production mechanism in SNe Ia, elements around the iron-peak are expected to follow Fe.
The [Cr/Fe] and [V/Fe] ratios in our stars are low (at $\sim -$0.28 to $-0.14$ dex).
We also find slightly subsolar values for 
[Ni/Fe], whereas the Co- and [Sc/Fe] ratios are slightly elevated and/or compatible with zero within the measurement errors. 

At a mean of $-0.43$ and $-0.70$ dex, the [Mn/Fe] and [Cu/Fe] ratios are very low, yet typical of metal poor halo field and GC stars. 
Overall, the Fe-peak elements are compatible with other GCs at similar 
metallicities and fall within the halo trends of, e.g., Cayrel et al. (2004) and Ishigaki et al. (2013); see also the discussions in Koch \& C\^ot\'e (2010).

Nissen \& Schuster (2011) reported on systematic differences in the abundance distribution of halo stars between 
$-1.6 < $[Fe/H] $ < -0.4$ in the Solar neighborhood, 
 which are most pronounced in Na, Cu, and Zn. This was extended into the metal poor regime ([Fe/H]$>-3.3$) by Ishigaki et al. (2013), confirming the 
 inner/outer, or, in-situ/accreted halo dichotomy in their Na/Fe and Zn/Fe ratios with the strongest signatures again occurring above $-1.5$ dex. 
 At the low metallicity of NGC~5897 we cannot unambiguously make such a distinction and, in a one-to-one comparison with the halo samples, 
 different elements would yield different associations with either halo component (e.g., a low [Cu/Fe] more in line with the low-$\alpha$, accreted 
 population while our elevated [Zn/Fe] ratio is rather reminiscent of the high-$\alpha$ halo stars that likely formed in-situ). 
\subsection{Neutron-capture elements: \hspace{5cm}Y, Ba, La, Ce, Pr, Nd, Sm, Eu, Gd}
We could reliably measure abundance ratios of nine chemical elements heavier than Z$>$38. 
As often stated for other elements above, also their mean values are rather normal and compatible with those of other GCs and field halo stars of the same metallicity (Pritzl et al. 2005; Tolstoy et al. 2009; Kacharov et al. 2013). 
In this regard, we measure an [$r$/$s$]-ratio, represented by [Eu/Ba], of 0.48 dex, which lies close to the Solar-scaled r-process value (e.g., Simmerer et al. 2004), as 
can be expected, since at these low metallicities AGB contributions were few and the bulk of the heavy elements was  
produced in the $r$-process (e.g, Truran 1981)\footnote{Note, however, that the $s$-process might have been active in the Galaxy as early as [Fe/H]$>-2.6$ (Simmerer et al. 2004).}. 


We note that none of our targets is exceptionally Ba-rich; the fact that also the intermediate-population, Na-rich stars have normal [Ba/Fe] ratios shows that they have not been enriched 
by great amounts of low-mass AGB ejecta that could have formed late in the central regions of the GC (cf. Kacharov et al. 2013). 

As Fig.~6 indicates, the mean heavy element abundances in NGC~5897 lie close to the Solar-scaled $r$-process curve (taken from Simmerer et al. 2004). 
An exception is Yttrium, which lies closer to the Solar $s$-process curve; Y, like its neighbour Sr, is associated with an additional, unknown, nucleosynthetic process, 
first identified by McWilliam (1998), and now often referred to as
Lighter Element Primary Process (LEPP; e.g. Travaglio et al. 2004).
\begin{figure}[htb]
\centering
\includegraphics[width=1\hsize]{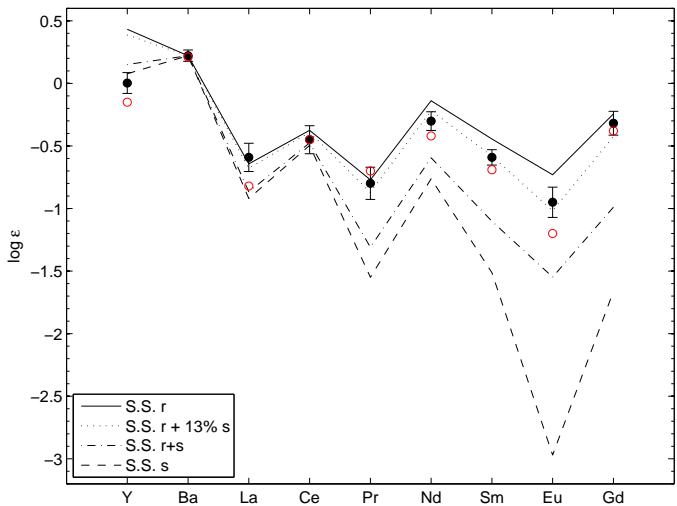}
\caption{Mean heavy element (Z$>$38) distribution  of the NGC~5897 stars. Different lines are the Solar-scaled (S.S.) abundances from Simmerer et al. (2004), normalized to Ba. We also indicate the 
best-fit to our measurements that is obtained by admixing 13\% of s-process material. The red open circles show the abundances for star S-138, which is low in Eu.}
\end{figure}
The best fit to the production of these heavy elements can be obtained for a Scaled-Solar pure r-process with an admixture of 
 some 13\% of s-process material (see also Kacharov et al. 2013). 
 This also accounts well for Zn, which is thought to have contributions from massive SNe II at very low metallicities as well 
 as nucleosynthesis in the weak $s$-process in massive stars, while low-mass AGB production appears to be negligible in this regime of NGC~5897 (e.g., Timmes  et al. 1995; Ishigaki et al. 2013). 
Other GCs show a clear-cut $r$-process dominance that remains even until higher metallicities (e.g., Pal 3, M15, M75; Koch et al. 2009; Sneden et al. 2000; Kacharov et al. 2013)
and has  been linked to the loose central concentration of those objects in the outer halo.  

The abundance mix we find in NGC~5897 at its low metallicity is fully consistent with the notion that only few AGB stars have contributed to the enrichment of the material, out of which the GC formed 
at early times. 
To emphasize this, we overplot in Fig.~7 the mean distribution of heavy elements in M15 by Sobeck et al. (2011), who determined abundance ratios of 22 neutron-capture elements in this metal-poor ([Fe/H]$\sim -2.6$) halo GC. 
The similarity of those elements measured in either study is obvious and clearly compatible with the negligible contributions by $s$-process nucleosynthesis. 
\begin{figure}[htb]
\centering
\includegraphics[width=1\hsize]{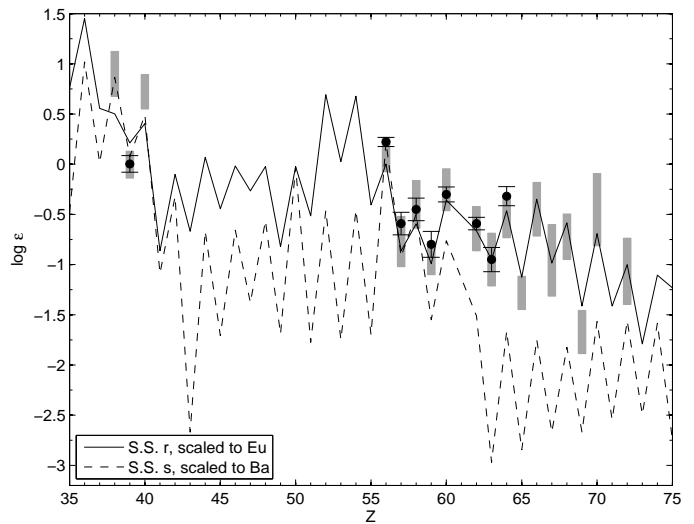}
\caption{Comparison of NGC~5897 (solid circles) with the heavy element abundances of M15 (Sobeck et al. 2011), where the gray shaded areas
frame their 1$\sigma$-ranges. In contrast to Fig.~6, the Solar-scaled curve for the $r$-process is shown relative to Eu, while the  $s$-process curve is still normalized to Ba.}
\end{figure}
\subsection{Notes on individual stars} 
Here, we comment on those stars amongst the, otherwise regular, sample, which show one or the other noteworthy abundance ratios.

{\em S-255:} This is the coolest and most metal-rich star of our sample. Some of its [$\alpha$/Fe] abundance ratios are  
significantly lower than the cluster mean (by $\sim$0.15 dex), in particular Si and Ca. 
Consequently, depletions are also found for K and Sc and some of the heavier elements (Y, La). 

{\em S-20/V-5}: The variable star candidate shows the {highest} Na-abundance in our sample, albeit at a regular O/Fe ratio. 
An association with the intermediate GC population is thus not unambiguous. 

{\em S-366}: This star is the best contender for a true second-generation star, given its very low O- and  very high Na-abundances.
None of its other element ratios are outstanding in any regard. 

{\em S-138:} With the highest O and a low Na-abundance, this star is a good representative of the primordial GC population. 
What distinguishes this object, however, are Eu and La abundances that are 
significantly lower than in the remainder of the GC stars analysed here (see Figures~6, 8),   
while Ba is higher than in the remainder of our sample. 
This is highlighted in Fig.~8, where we compare two  absorption features for Ba and Eu used in our abundance analysis for the the Eu-poorer S-138 
and a typical cluster representative, S-290\footnote{A similar, obvious depletion is seen for the stronger Eu~{\sc ii}-line at 4129\AA, which is, however, 
hampered by difficulties in the continuum setting due to blending,  and thus was not used in our abundance analysis.}. 
\begin{figure}[htb]
\centering
\includegraphics[width=1\hsize]{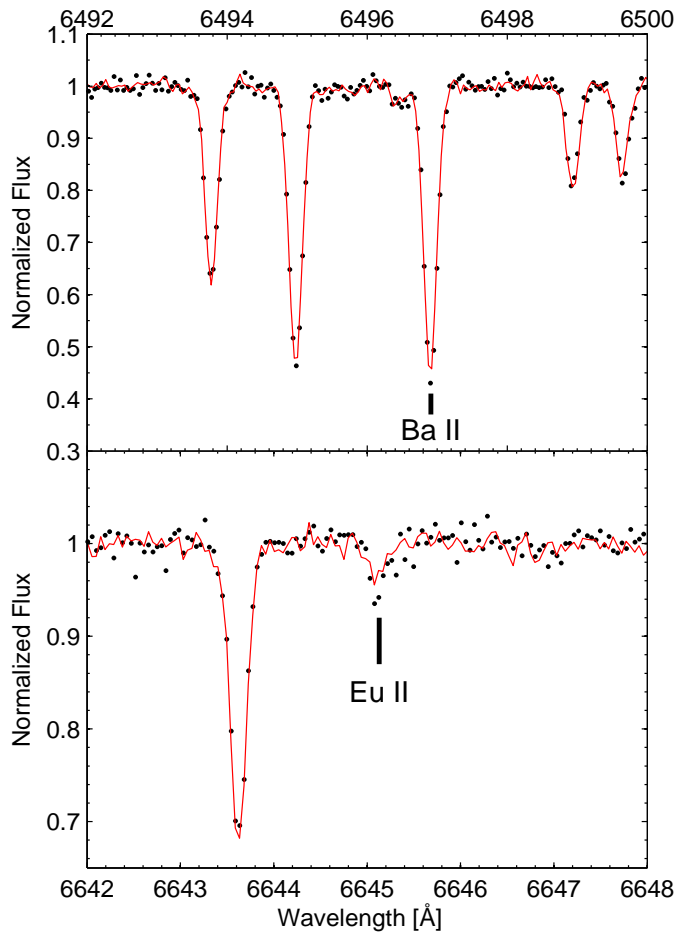}
\caption{Comparison of the Ba~{\sc ii} 6496.9\AA-line (top panel) and Eu~{\sc ii} 6645\AA-line (bottom panel) in two stars with similar stellar parameters: 
S-290 (black dots) and the primordial star S-138 (red solid line), which shows low Na, La, Eu, and Nd abundances.}
\end{figure}
{In fact, Eu can be depleted very easily due to its large neutron-capture cross section, similar to La, which contains 
a larger $r$-process fraction relative to Ba. }

Accordingly, S-138 stands out in that it has a [Ba/Eu] that is higher by 0.28 dex than the other six red giants. 
On the other hand, its [La/Y]  (thus a [$hs$/$ls$]) ratio of 0.41 is consistent with the GC's mean value, indicating that this star was not polluted by an unusual mass range of progenitors, 
but rather the same sources (e.g., low-intermediate mass, metal poor AGB stars; Cristallo et al. 2009) produced the observed abundance patterns. 
\section{Summary}
Our abundance analysis of a large number of chemical tracers in the metal-poor inner halo cluster NGC~5897 identified
this GC as fully representative of the bulk of such objects in the Milky Way. 
This includes the enhancement in the $\alpha$-elements to $\sim$0.3 dex, approximately Solar Fe-peak elements, 
and n-capture elements that were predominantly produced by the $r$-process. The later notion is consistent with only negligible AGB contributions
to the chemical inventory in the metal poor (read: early-time) regime. 
The presence of a pronounced Na-O anti-correlation renders this object, by definition, a {\em globular} cluster (not that 
there ever was a doubt). 
Unfortunately, our limited sample size did not allow us to trace the full extent of this relation, in particular we are lacking 
information about the spatial variations of the first vs. second generations, but the severe crowding prohibited exposures of the central stars. 
All other elements, safe for O, Na, and Al only show little star-to-star scatter, mostly explicable by the measurement uncertainties or 
in part driven by peculiar patterns in individual stars. 

Several features such as the shape of the Na-O correlation and the preponderance of the $r$-process is reminiscent of the metal poor ([Fe/H]$\sim -2.6$ dex) halo GC M15.

Our work has revealed NGC~5897 to be more metal poor than implied by older CMD and low-resolution metallicity studies. 
We suggest that this lower metallicity can explain the remarkably longer periods of the RR Lyr found in this GC. 
Indeed, the error-weighted mean metallicity of the RR Lyr of Clement \& Rowe (2001) is [Fe/H] = $-2.2\pm0.2$ on the scale of Zinn \& West (1984), where we adopted the relations between period, phase, and metallicity 
 of Jurcsik (1995);  see also Haschke et al. (2012). 
While the errors on individual stars are tremendous, the mean value of the RR\,Lyr is thus in line with the metal poor nature of this GC found from our red giants. 
\begin{acknowledgements}
AK acknowledges the Deutsche Forschungsgemeinschaft for funding from  Emmy-Noether grant  Ko 4161/1. 
We thank the anonymous referee for a helpful and constructive report.
This research made use of atomic  data from the INSPECT database, version 1.0 (www.inspect-stars.net). 
The CSS survey is funded by the National Aeronautics and Space
Administration under Grant No. NNG05GF22G issued through the Science
Mission Directorate Near-Earth Objects Observations Program.  The CRTS
survey is supported by the U.S.~National Science Foundation under
grants AST-0909182 and AST-1313422.%
 \end{acknowledgements}

\end{document}